\documentclass[fleqn,usenatbib]{mnras}

\usepackage{newtxtext,newtxmath}

\usepackage[T1]{fontenc}

\DeclareRobustCommand{\VAN}[3]{#2}
\let\VANthebibliography\thebibliography
\def\thebibliography{\DeclareRobustCommand{\VAN}[3]{##3}\VANthebibliography}

\usepackage{graphicx}	
\usepackage{amsmath}	
\usepackage{rotating}
\usepackage{pdflscape}
\usepackage{subfig}
\usepackage{hyperref}
\usepackage{tablefootnote}
\usepackage{threeparttable}
\usepackage{bbding}
\usepackage{booktabs}
\usepackage{longtable}
\DeclareSymbolFont{matha}{OML}{txmi}{m}{it}
\DeclareMathSymbol{\varv}{\mathord}{matha}{118}
\usepackage[inline]{enumitem}

\title[SPHEREx Wolf-Rayet stars]{Detection of new Galactic Wolf-Rayet stars with SPHEREx}

\author[Paul A. Crowther \& Erin L. Smith]{
Paul A. Crowther$^{1}$\thanks{E-mail: Paul.Crowther@sheffield.ac.uk},
Erin L. Smith$^{1}$
\\
$^{1}$Astrophysics Research Cluster, School of Mathematical and Physical Sciences, University of Sheffield, Hicks Building, Hounsfield Road, Sheffield S3 7RH, UK
}

\date{Accepted XXX. Received YYY; in original form ZZZ}

\pubyear{\the\year{2026}}

\begin{document}
\label{firstpage}
\pagerange{\pageref{firstpage}--\pageref{lastpage}}
\maketitle


\begin{abstract}
We assess the potential for the NASA Explorer SPHEREx mission in adding to the Milky Way census of Wolf-Rayet (WR) stars by extracting 0.75--5$\mu$m spectrophotometry of 30 candidate WR stars previously identified using
{\it Gaia} XP spectra. We confirm an additional Galactic 13 WR (8 WN and 5 WC) stars, identify 3 probable [WC]-type central stars of Planetary Nebulae, plus 2 He emission line sources. 
SPHEREx has the potential to detect additional relatively isolated WR stars on the far size of the Milky Way by combining Machine Learning techniques with a judicious choice of candidate selection criteria. Indeed SPHEREx is
sufficiently sensitive to identify bright  Magellanic Cloud Wolf-Rayet stars.
\end{abstract}




\section{Introduction}

Classical Wolf-Rayet (WR) stars are evolved descendents of massive OB-type stars, possess dense, fast outflows and so exhibit unusually strong, broad emission lines across the electromagnetic spectrum \citep{Shenar2026}. Their strong winds often
dictate the fate of  massive stars \citep{Romagnolo+2024}. Very massive stars (VMS) also resemble WR stars owing to their proximity to the Eddington limit \citep{Crowther+2010}. Consequently, WR stars and VMS represent useful tracers of recent star-formation in galaxies, and provide sources of mechanical, radiative and chemical feedback in young
populations \citep{Crowther2007}. A subset of H-depleted planetary nebulae central stars also exhibit similar spectral morphologies \citep{TodtHamann2015, Weidmann+2020}.

Historically, Galactic WR stars were primarily detected serendipitously via  H$\alpha$ surveys \citep[e.g.][]{Hopewell+2005} or dedicated optical narrow-band surveys \citep{Shara+1999}.
 Since they are primarily located in the dusty thin-disk of the Milky Way, detected sources are generally confined to the Solar Neighbourhood, such that the Galactic WR census stood at 227 a quarter of a century ago \citep{vanderHucht2001}.
Fortunately, the advent of efficient large format infrared detectors over the last couple of decades has permitted searches for Wolf-Rayet stars and other emission line sources within the entire Galactic plane \citep{Crowther2015}. To date, approximately 700 Wolf-Rayet stars have been identified in the Milky Way\footnote{v1.35 of the online catalogue maintained at \url{http://pacrowther.staff.shef.ac.uk/WRcat}}, yet the total has been estimated to lie between 1500 \citep{Shara+1999} and 6500 \citep{vanderHucht2001}. 

In particular, since the infrared colours of Wolf-Rayet stars are unusual, owing to free-free emission from their dense stellar winds, candidates have been identified from colour-colour diagrams \citep{Hadfield+2007, Mauerhan+2011, Faherty+2014}, combining Two Micron All Sky Survey \citep[2MASS,][]{2MASS} with {\it Spitzer}/GLIMPSE \citep{Benjamin+2003} or {\it WISE} \citep{WISE} observations of the Galactic plane. In parallel,  \citet{Shara+2009, Shara+2012} used several narrow K$_{s}$-band filters -- sensitive to specific WR emission lines -- to image the Galactic plane using ground-based telescopes, resulting in complementary Wolf-Rayet candidates. Nevertheless, such searches are relatively inefficient since WR stars are rare, and contaminants are common \citep[e.g.][]{Hadfield+2007, RossloweCrowther2018}. Rich, dust-obscured, young star clusters have also been successfully targeted \citep{Crowther+2006, Davies+2012, Chene+2013}. 

The bulk of available infrared WR spectra is restricted to the JHK windows between 1--2.5$\mu$m \citep{Figer+1997, Crowther+2006, Nishimaki+2008, Shara+2012}, although spectroscopy beyond 2.5$\mu$m has been obtained with the SpeX instrument at the {\it Infra Red Telescope Facility} \citep{Faherty+2014}, plus various space-based telescopes, including the Short-Wavelength Spectrometer (SWS)  aboard {\it Infrared Space Observatory} \citep{vanderHucht+1996, Dessart+2000, Morris+2000}, the Infrared Spectrograph (IRS) on {\it Spitzer} \citep{Morris+2004, Ignace+2007}, Photodetector Array Camera and Spectrometer (PACS) aboard {\it Herschel} \citep{Crowther+2024}, plus the Mid Infrared Instrument (MIRI) on {\it JWST} \citep{Lau+2022}.

In this study we combine Spectro-Photometer for the History of the Universe, Epoch of Reionization, and Ices Explorer  \citep[SPHEREx,][]{Bock+2026} with candidate WR stars identified from a Machine Learning analysis \citep{Creevey+2023} of low-resolution {\it Gaia} BP/GP (hereafter XP) spectroscopy \citep{DeAngeli+2023}. A subset of such WR candidates have been  confirmed via follow-up optical spectroscopy \citep{Marin+2024, Mulato+2025}, but many additional candidates exist, representing test cases for the capabilities of SPHEREx for Milky Way emission line stars. Machine learning techniques sometimes identify emission line sources with some spectroscopic similarities to WR stars, albeit with a very different nature such as Cataclysmic Variables \citep[e.g. SW~Sex,][]{Chang+2025}.
{\it Gaia} XP candidates and SPHEREx are introduced in Section~\ref{obs}, low resolution spectroscopy of WR stars and related sources are
presented in Section~\ref{old}, new candidates are discussed in Section~\ref{new}, with brief discussion and conclusions drawn in Section~\ref{conc}.

\section{Spectrophotometry}\label{obs}

\begin{figure*}
    \centering
  \includegraphics[width=2.0\columnwidth]{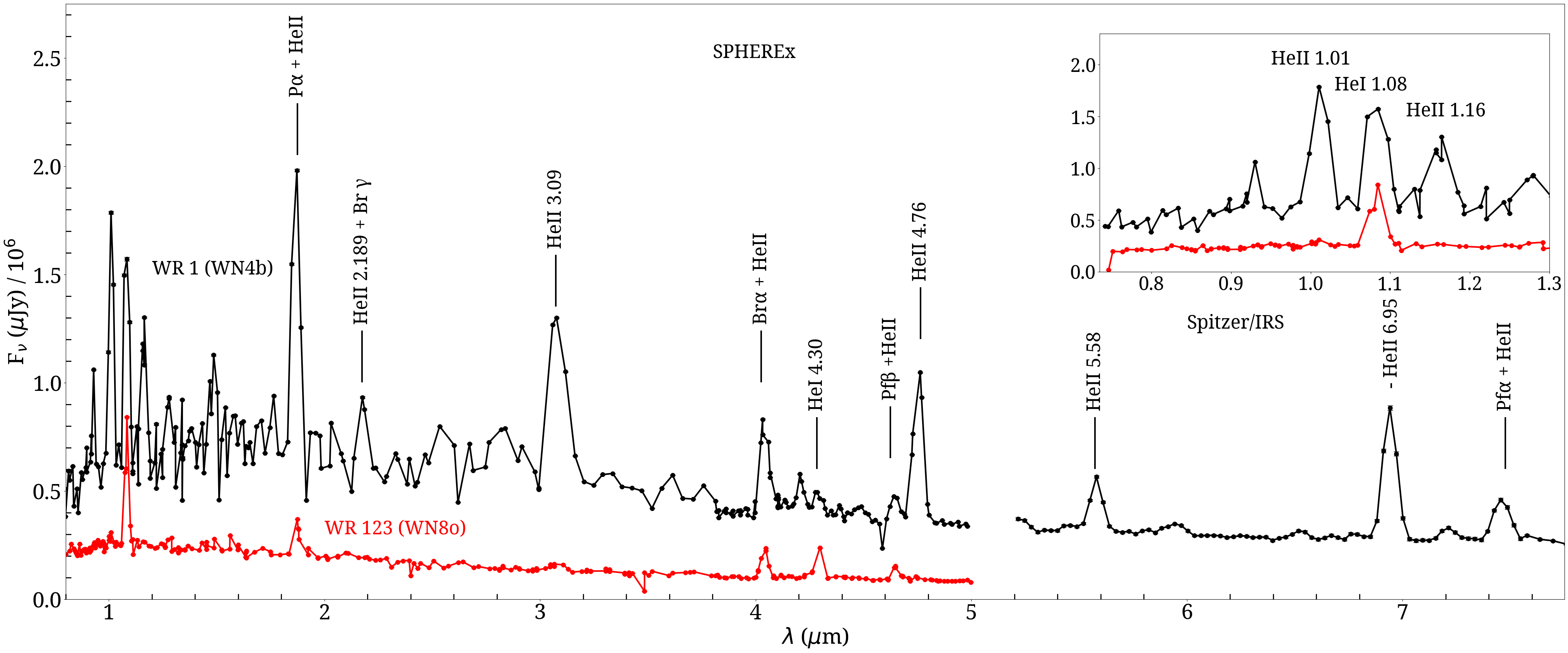}
   \includegraphics[width=2.0\columnwidth]{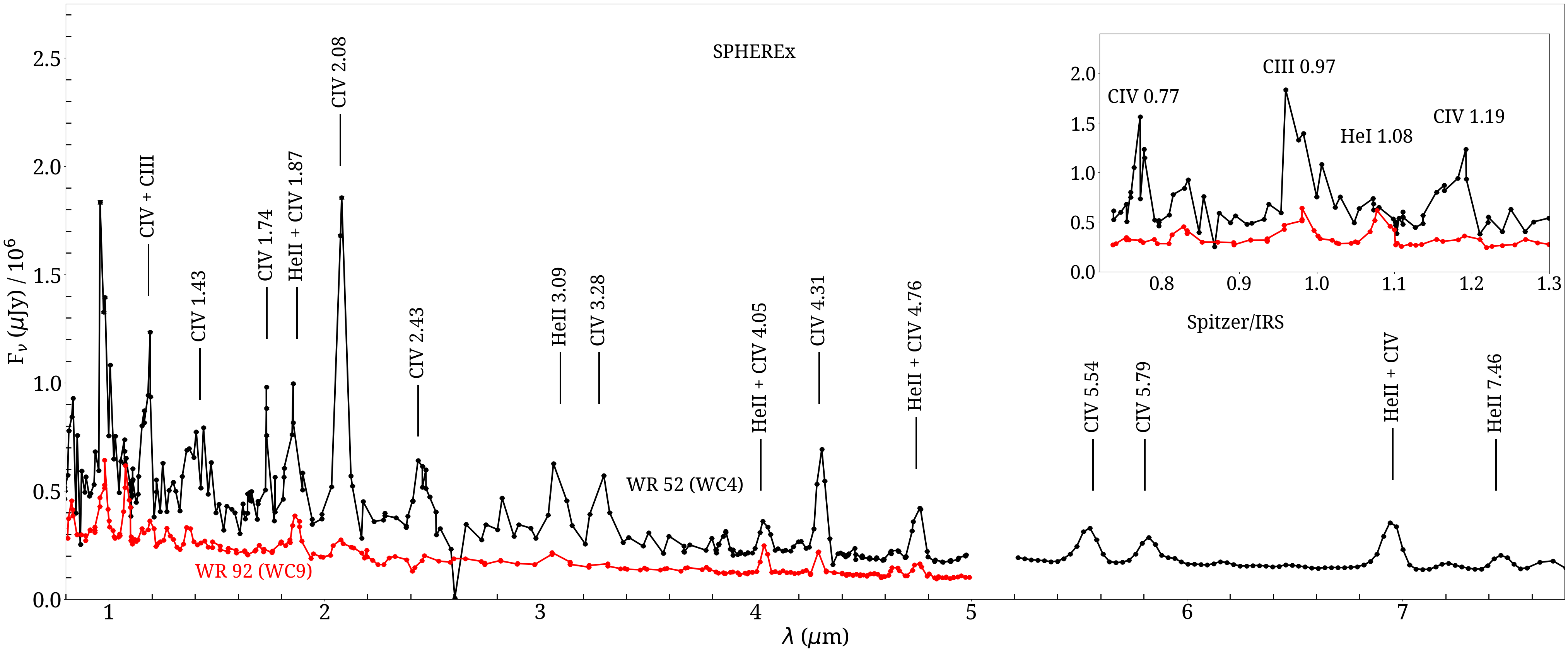}
  \includegraphics[width=2.0\columnwidth]{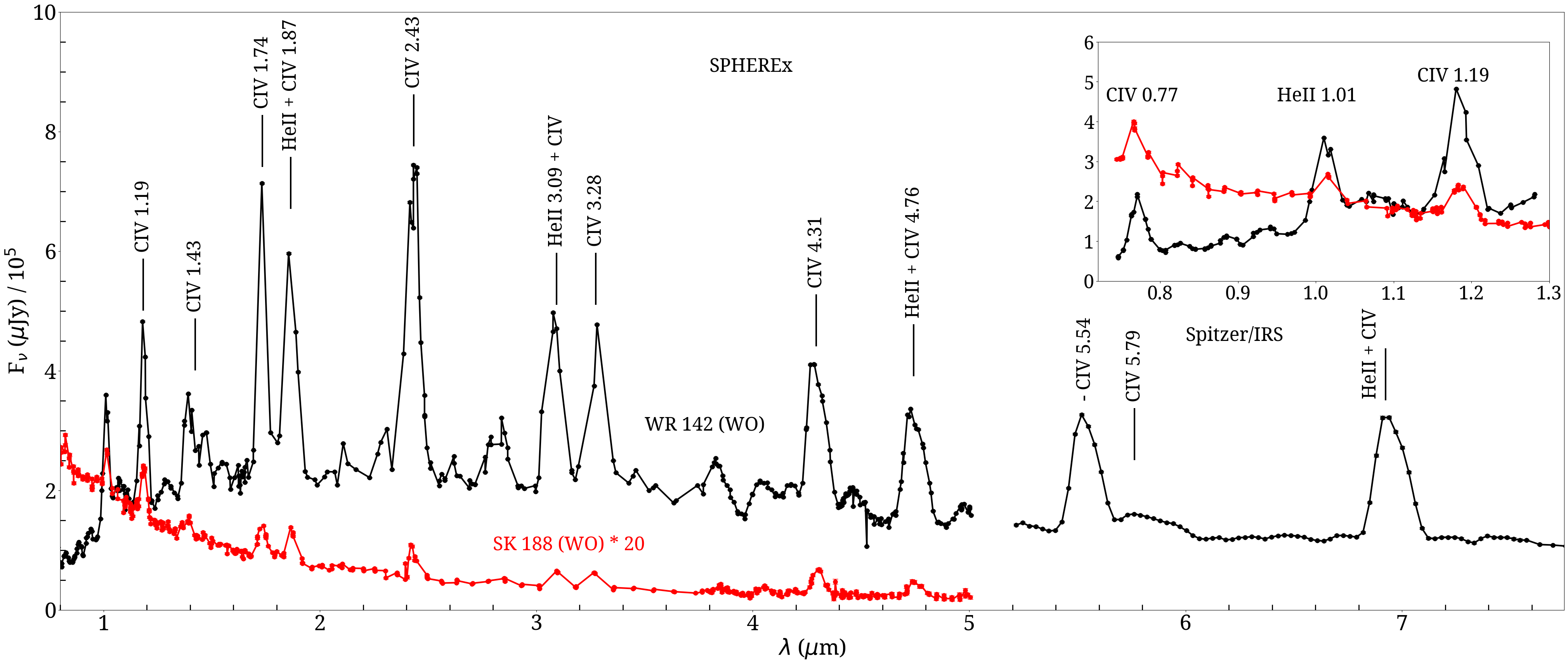}
         \caption{SPHEREx spectrophotometry of representative  WN stars WR1 (black) and WR123 (red) in the upper panel, WC stars WR52 (black) and WR92 (red) in the middle panel and the WO stars WR142 (black) and Sk 188 (red, SMC star whose flux has been scaled by factor of 20) in the lower panel. Low resolution (SL) {\it Spitzer}/IRS spectroscopy of WR1, WR52 and WR142 is included, with zoomed in spectra around 1$\mu$m shown in the upper right side of each panel.}
         \label{SPHEREx-WR}
    \end{figure*}

\subsection{SPHEREx}

SPHEREx \citep{Bock+2026} is a NASA Explorer mission undertaking an all-sky survey across 102 narrow-bands between 0.75--5$\mu$m, corresponding to a resolving power of $R$ =  35-41 ($\leq$ 3.8$\mu$m, Bands B1-B4), $R$ = 110 (3.8-4.4$\mu$m, Band B5) or $R$ = 130 (4.4-5.0$\mu$m, Band B6) at a coarse spatial resolution (6.2$''$ pixels). SPHEREx was launched on 11 March 2025, commenced regular science operations on 1 May 2025 and is due to complete 4 observations of the sky in each filter position over its nominal 2 year operation. The SPHEREx data processing pipeline is described in \citet{SPHEREx-pipeline}. The SPHEREx Spectrophotometry Tool\footnote{\url{https://irsa.ipac.caltech.edu/applications/spherex/tool-spectrophotometry}} provides forced photometry of point sources for each narrow-band image, with a choice of background region, defaulting to  15 pixels, using {\tt the Tractor}\footnote{\url{http://github.com/dstndstn/tractor}}. Calculation of the background incorporates catalogued IR 
point sources, or these can be manually added when undertaking spectrophotometry of targets. Observations discussed here were processed in July 2026, so generally represent two observations per filter position.

In principle, the experimental design of SPHEREx is well suited to the identification of heavily reddened WR stars owing to their rich emission line spectra and their predicted peak K$_{s}$-band magnitude of 
$\sim$10  \citep{RossloweCrowther2015}, which corresponds to $F_{\nu} = 65$ mJy. Indeed, the majority of all Galactic WR stars are anticipated to be brighter than $K_{s}$ = 15 mag or $F_{\nu} = 0.65$ mJy \citep[][their figure~12]{RossloweCrowther2015} for which solely strong emission lines would readily be detected. In addition, the brightest WR stars at IR wavelengths may suffer from saturation effects.
In summary, SPHEREx covers a similar spectral range to {\it JWST} NIRSpec/PRISM \citep{NIRSpec}, albeit with a somewhat lower resolving power, and is suited to milli-Jy  rather than nano-Jy sources.

\begin{table}
    \caption{Low resolution (SL, 5.2--14.5$\mu$m) {\it Spitzer}/IRS datasets of WR stars and related emission line sources used in the present study, including AOR (Astronomical Observation Request) number. {\it Spitzer}/IRS datasets of WR1 and WR52 have previously been discussed by \citet[][]{Ignace+2007}.}
    \centering
    \label{IRS}
    \begin{tabular}{
l@{\hspace{2mm}}
l@{\hspace{2mm}}
l@{\hspace{2mm}}
l@{\hspace{2mm}}
l}
    \hline
    Star & Sp Type & Obs & AOR & PI (programme) \\
 \hline
    WR1 & WN4b & Oct 2004 & 10440704 & Leitherer (3185) \\
    WR52 & WC4 & Mar 2006 & 14237440 & Crowther (20154)  \\
    WR142 & WO2  & Jun 2005 & 12414464 & Houck (199) \\
    NGC~6905 & [WO1] & Jun 2007 & 21969152 & Fazio (40115) \\
    NGC~40 & [WC8] & Nov 2007 & 21976832 & Fazio (40115) \\
    M4-18 & [WC10] & Nov 2008 & 21957632 & Bernard-Salas (40035) \\
    He~3-519 & early B\,Ia$^{+}$ & May 2005 & 13735680 & Calibration \\
     \hline
    \end{tabular}
    \end{table}

 \subsection{ {\it Spitzer}/IRS}\label{Spitzer_IRS}

Although the primary focus of this study is SPHEREx, a subset of Wolf-Rayet stars and related emission line objects have been observed with the Infrared Spectrograph (IRS) aboard {\it Spitzer} \citep{Houck+2004}. The bluest
module (Short-Low, SL) of IRS covered the 5.2--14.5$\mu$m range at a resolving power of $\sim$100, in first (7.6--14.5$\mu$m, SL1) or second (5.2--7.6$\mu$m, SL2) order, with a 2 pixel slit width corresponding to 3.6$''$ 

Therefore, {\it Spitzer}/IRS provides spectral coverage of WR stars adjacent to SPHEREx at a comparable resolving power. Previous studies of WR stars using
{\it Spitzer}/IRS have included \citet{Morris+2004} and \citet{Ignace+2007}, which
primarily focused on mid-IR fine structure lines. Archival SL {\it Spitzer}/IRS spectroscopy retrieved from the {\it Spitzer} Heritage Archive were reduced using the final version (v18.18) of the pipeline\footnote{\url{https://www.ipac.caltech.edu/doi/irsa/10.26131/IRSA399}}. Table~\ref{IRS} provides details of IRS spectroscopy presented in this study, including Astronomical Observation Request (AOR) numbers, since some targets have multiple observations in the archive. All datasets used were acquired in Stare (as opposed to Map) mode.

\begin{figure*}
    \centering
   \includegraphics[width=2.0\columnwidth]{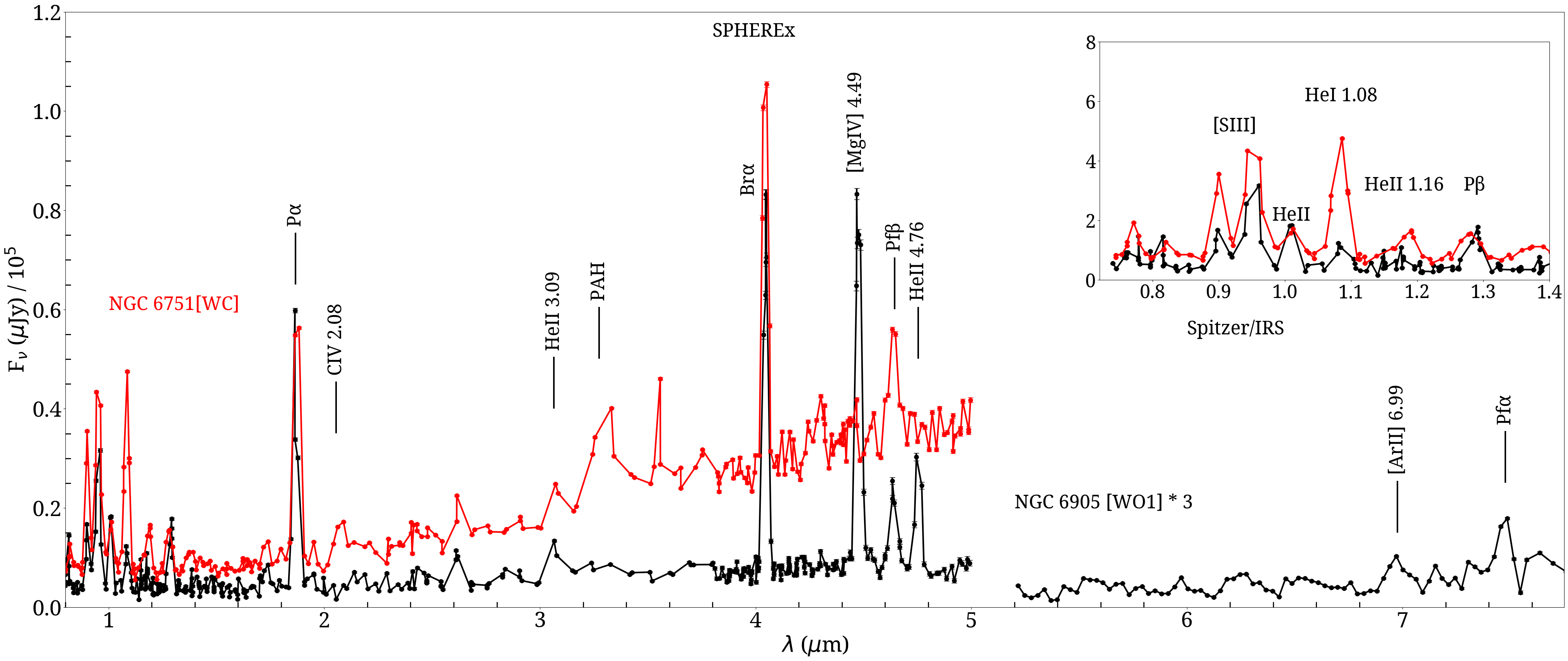}
   \includegraphics[width=2.0\columnwidth]{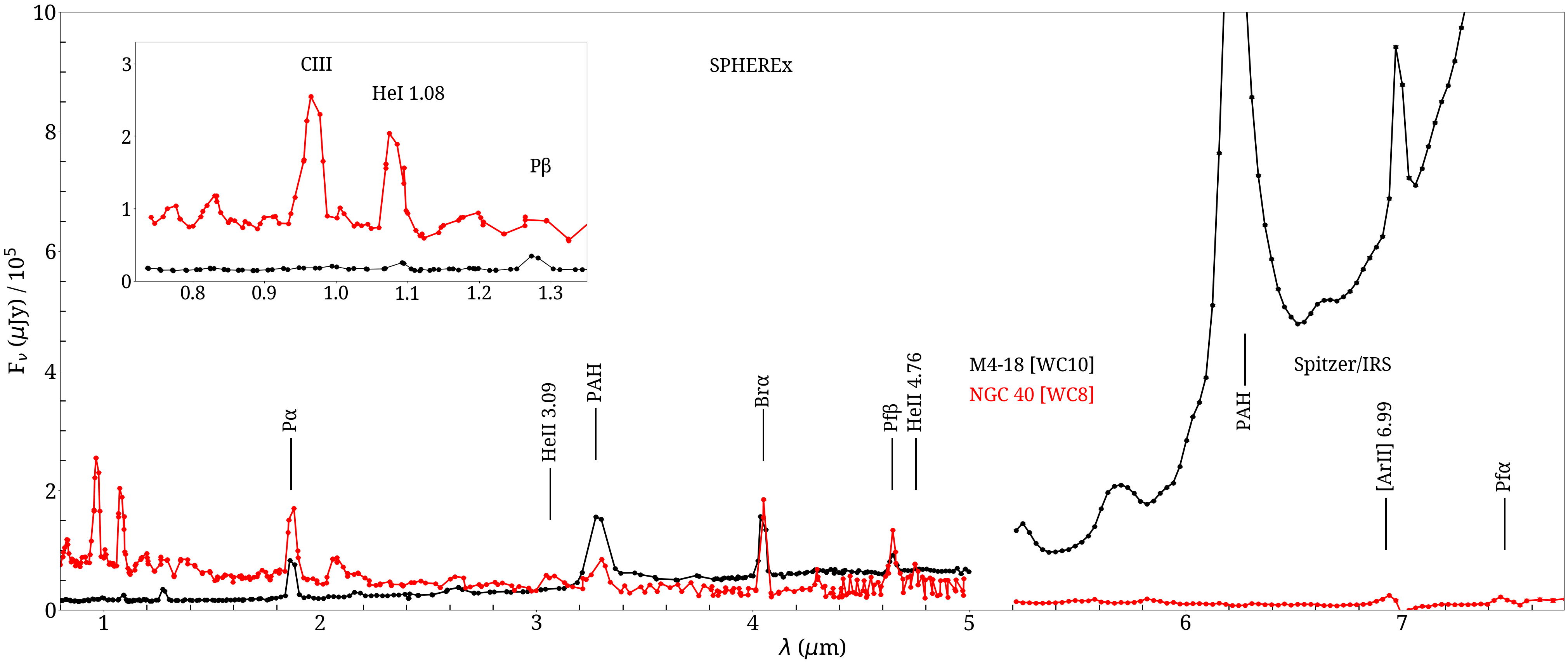}
  \includegraphics[width=2.0\columnwidth]{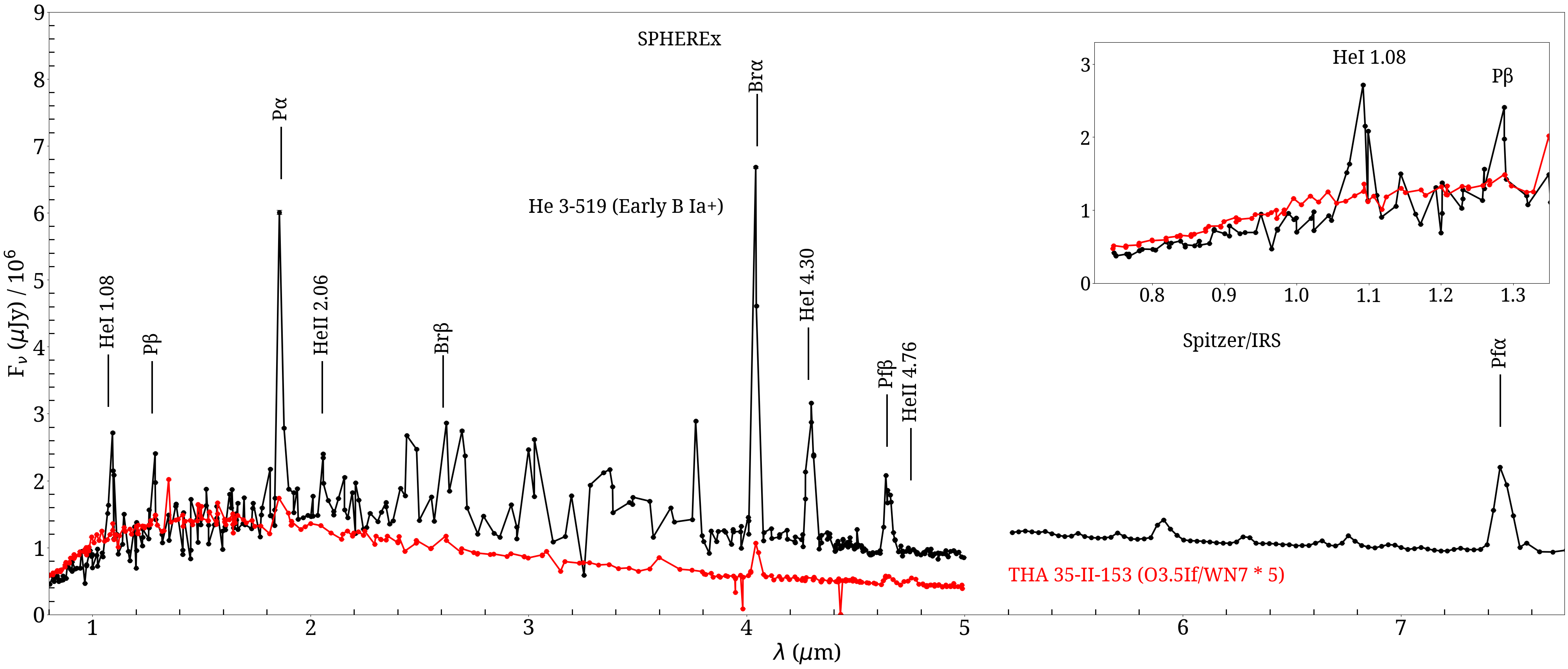}
         \caption{SPHEREx spectrophotometry of high excitation Wolf-Rayet
         type CSPNe NGC~6905 (black, [WO]) and NGC~6751 (red, [WC4]) in the upper panel, low excitation CSPNe M4-18 [(black, [WC10]) and NGC~40 (red, [WC8]) in the middle panel and extreme OB supergiants He~3--519 (early B supergiant, black) and THA~35-II-153 (O3.5\,If/WN7, red) in the lower panel. Low resolution (SL) {\it Spitzer}/IRS spectroscopy of NGC~6905, M4-18, NGC~40 and He~3--519 is included, revealing strong PAH emission at 6.2$\mu$m in M4-18, with zoomed in spectra around 1$\mu$m shown in each panel. The flux of NGC~6905 has been scaled by a factor of 3 and THA~35-II-153 (WR31-1) has been scaled by a factor of 5 for clarity.}
         \label{cspn_obsuper}
    \end{figure*}

    \subsection{Emission line stars from {\it Gaia} XP}

Beyond its astrometric capabilities, {\it Gaia} also provides medium resolution spectroscopy of visually bright sources around the calcium triplet \citep[RVS,][]{rvs}, plus low resolution XP spectrophotometry for sources with $G<$ 19 mag \citep{DeAngeli+2023}. 
The latter XP dataset was released in Data Release 3, covering the wavelength ranges 0.33-0.68~$\mu$m (BP, $R$=30-100) and 0.64-1.05$\mu$m (RP, $R$=70-100), and correspond to cut-outs of 3.5$\times$2.1 arcsec$^{2}$ \citep{Carrasco+2021}. \citet{Creevey+2023} introduce a Machine Learning approach in which the Extended Stellar Parametrizer for Emission Line Stars (ESP-ELS) module assigns probabilities for Be, Herbig Ae/Be, T Tauri, active M dwarfs, Planetary Nebulae  (PNe) or Wolf-Rayet stars based on their {\it Gaia} XP spectra. \citet{Mulato+2026} discuss candidate PNe from ESP-ELS.

Amongst the 565 Wolf-Rayet candidates from ESP-ELS, half have an assigned
{\tt classlabel$\_$espels$\_$flag} of 0 (very high probability), with the remainder spread between flags 1 (high probability) to 4 (low probability). The overwhelming majority of sources with a 0 flag are previously known Milky Way or Magellanic Cloud WR stars. \citet{Mulato+2025} identified 59 candidate WR stars from ESP-ELS, utilising IR broad-band colour criteria of \citet{Mauerhan+2011}. Of these, the brightest 37 candidates were observed spectroscopically, with 33 sources confirmed - three of which were in common with \citet{Marin+2024}. We shall investigate the nature of the remaining high probability ESP-ELS candidate WR stars. All candidates discussed, with the exception of the Planetary Nebula NGC~6751 and candidate [WC]-type star G326.5-0.6, have {\it Gaia} re-normalised unit weight error (RUWE) values below the 1.25 threshold recommended to exclude unresolved binaries \citep{RUWE}.

\section{SPHEREx observations of WR stars and related objects}\label{old}

\subsection{Wolf-Rayet stars}

Representative spectrophotometry of (mostly) Galactic Wolf-Rayet stars from SPHEREx are presented in Fig.~\ref{SPHEREx-WR}, including archival {\it Spitzer}/IRS low-resolution spectroscopy, where available, which extend to Pf$\alpha$ (Sect.~\ref{Spitzer_IRS}). Owing to the low resolving power of SPHEREx, we have also consulted unpublished {\it IRTF}/SpeX observations of Galactic WR stars (W.D.~Vacca, priv. comm.), which helped establish the
near-IR classification scheme for WR stars of \citet{RossloweCrowther2018}. 

The upper panel compares the strong-lined early WN star WR1 (HD~4004, black) to the late WN star WR123 (HD~177230, red), including a zoomed in view of the critical 1$\mu$m region in the top right subpanel. The most prominent emission lines in WR1 are He\,{\sc ii} 1.01$\mu$m (5--4), 1.86$\mu$m (6--5), 3.09$\mu$m (7--6), 4.76$\mu$m (8-7), 6.95$\mu$m (9-8, {\it Spitzer}/IRS), plus He\,{\sc i} 1.08$\mu$m. WR 1
is hydrogen-free \citep{Hamann+2006}, so leading hydrogen lines are produced by corresponding He\,{\sc ii} lines, such as He\,{\sc ii} (10-8) instead of Br$\alpha$ (5--4). Unusually for late WN stars, WR123 
is also highly deficient in hydrogen 
\citep{MasseyConti1980, Crowther+1995}, and since He\,{\sc ii} lines are weakly present in late WN stars, He\,{\sc i} components of (4--3) and (5--4) transitions will dominate the prominent P$\alpha$ 
and Br$\alpha$ features \citep{Morris+2000} as well as He\,{\sc i} 1.08$\mu$m and 4.30$\mu$m. 
Weaker near-IR He\,{\sc i} lines at 1.70$\mu$m and 2.06$\mu$m are difficult to identify with SPHEREx. 

The middle panel of Fig.~\ref{SPHEREx-WR} compares the early WC star WR52 (HD~115473, black) to the WC9 star WR92 (HD~157451, red). The most prominent emission lines in WR52 are C\,{\sc iv} 2.08$\mu$m, followed by C\,{\sc iv} $\alpha$ transitions, including 1.19$\mu$m (8--7), 1.74$\mu$m (9--8), 2.43$\mu$m (10--9), 3.28$\mu$m (11--10), 4.31$\mu$m (12--11), 5.54$\mu$m (13--12, {\it Spitzer}/IRS), plus C\,{\sc iii} 0.97$\mu$m and He\,{\sc ii} 1.01$\mu$m, 3.09$\mu$m, 4.76$\mu$m. The high ionization features are less pronounced in the WC9 star, with C\,{\sc iii} 0.97$\mu$m and He\,{\sc i} 1.08$\mu$m amongst the most prominent features (see zoomed in view in the subpanel).

The lower panel of Fig.~\ref{SPHEREx-WR} shows the WO stars WR142 (Sand 5, black) and Sk ~188 (AB8, SMC, red) whose IR spectra are dominated by most of the same C\,{\sc iv} emission lines as WR52, 
with the exception of 2.08$\mu$m which is suppressed in WO stars \citep[see][]{Tramper+2015}, plus corresponding He\,{\sc ii} transitions, with He\,{\sc i} absent. C\,{\sc iv} 0.77$\mu$m and 1.19$\mu$m are clearly observed in WO stars (see zoomed in subpanel), aided by the high wind velocities of WO stars \citep{KingsburghBarlow1995}. The SMC WO binary is relatively faint (K$_{\rm s}$ = 13.3 mag) so its flux has been scaled by a factor of 20. Indeed, Sk~188 \citep[WO+O4,][]{Moffat+1985} is one of numerous Magellanic Cloud WR systems to be readily detected with SPHEREx.

Emission lines in dusty WC stars will often be masked by hot dust \citep{Williams2019}, except at the bluest SPHEREx wavelengths (I, Y and J bands), and the low spectral resolution will hinder the detection of weak emission lines, such as He\,{\sc ii} 1.01$\mu$m, 3.09$\mu$m in late WN stars (e.g. WR123 in Fig.~\ref{SPHEREx-WR}). The coarse spatial resolution of SPHEREx may also lead to severe dilution of WR emission lines by nearby IR continuum sources. This can be assessed by comparison with higher spatial resolution ($\sim2''$) photometry from 2MASS \citep{2MASS} and/or the GLIMPSE survey using {\it Spitzer}/IRAC \citep{Benjamin+2003}. {\it WISE} W1 (3.5$\mu$m) and W2 (4.6$\mu$m) band observations provide a superior sky coverage to GLIMPSE  \citep{WISE}, albeit at comparable spatial resolution to SPHEREx.

\subsection{Ionized nebulae}

The low resolving power of SPHEREx generally prevents strong nebular emission lines (tens of km\,s$^{-1}$)  being easily distinguished from Wolf-Rayet emission lines (thousands of km\,s$^{-1}$), since $R \sim$ 40 for Bands B1 to B4 corresponds to FWHM$\sim$ 7500 km\,s$^{-1}$. 
Exceptions include WO stars  (e.g. WR142 in Fig.~\ref{SPHEREx-WR}) and broad spectral lines in the highest resolution Band~B6 for which FWHM$\sim$ 2300 km\,s$^{-1}$. H\,{\sc ii} regions will generally exhibit strong nebular P $\alpha$, Br $\alpha$ and He\,{\sc i} emission in the SPHEREx wavelength range, together with  [S\,{\sc iii}] 0.91+0.95$\mu$m, but their ionizing sources are insufficient hard to produce 
nebular He\,{\sc ii} 1.01, 1.86, 3.09 or 4.76$\mu$m, and often show broad 3.3$\mu$m PAH emission in their associated photo dissociation regions \citep[PDRs,][]{Verstraete+2001}. 

In contrast, the ionizing stars of Planetary Nebulae (PNe) are capable of producing nebular He\,{\sc ii} in addition to H\,{\sc i}, He\,{\sc i}, plus forbidden  [S\,{\sc iii}] 0.91+0.95$\mu$m as well as [Mg\,{\sc iv}] 4.49$\mu$m \citep[][]{BernardSalas+2001}, so
the presence of collisionally excited forbidden lines helps PNe to be distinguished from WR stars.  By way of example, Fig.~\ref{CSPNe} includes SPHEREx spectrophotometry of the planetary nebula 
NGC~2392 whose central star (CSPN) is a subdwarf O star \citep{Pauldrach+2004, Kudritzki+2006}.

PNe with [WC] or [WO]-type central stars are more problematic regarding the potential for confusion with WC and WO stars, since they also 
produce stellar He\,{\sc ii} and C\,{\sc iv} emission. To illustrate their similarities, the upper panel of Fig.~\ref{cspn_obsuper} presents SPHEREx spectrophotometry of high excitation CSPNe, NGC~6905 and NGC~6751, which have [WO1] and [WC4] spectral types, respectively \citep{Crowther+1998}, together with {\it Spitzer}/IRS spectroscopy of NGC~6905. C\,{\sc iv} 2.08$\mu$m is observed in NGC~6751, while   other leading stellar C\,{\sc iv} recombination lines (e.g. 1.74$\mu$m) are present in both sources, suggesting a mix of stellar and nebular origin for He\,{\sc ii} 1.01$\mu$m, 1.86$\mu$m,  3.09$\mu$m, 4.76$\mu$m emission. Both exhibit nebular  [S\,{\sc iii}] 0.91+0.95$\mu$m (see zoomed in subpanel), plus [Mg\,{\sc iv}] 4.49$\mu$m is observed in NGC~6905.

The middle panel of Fig.~\ref{cspn_obsuper} presents SPHEREx spectrophotometry of low excitation CSPNe, M4-18 and NGC~40, which have [WC10] \citep[][]{DeMarcoCrowther1999} and [WC8] \citep[][]{Crowther+1998} spectral types, respectively, together with {\it Spitzer}/IRS spectroscopy. M4-18 is dominated by nebular emission, including [Ar\,{\sc ii}] 6.99$\mu$m plus strong PAH emission (6.2$\mu$m, 3.3$\mu$m), plus Paschen and Brackett lines and He\,{\sc i} 1.08$\mu$m (see zoomed in subpanel). NGC~40 emission includes nebular Paschen and Brackett lines, plus stellar wind lines (e.g. C\,{\sc iii} 0.97$\mu$m in subpanel) from the [WC 8] central star, so would be difficult to distinguish from WC stars solely from IR spectroscopy.


\subsection{OB supergiants}

It has been established for several decades that early-type supergiants exhibit IR emission lines \citep{McGregor+1988, Hanson+1996, Mauerhan+2010, Clark+2012}. In particular, very strong H\,{\sc i} and He\,{\sc i} emission lines are observed in B hypergiants such as P Cygni \citep{Najarro+1997}, HDE~316285 \citep{Hillier+1998} and He~3-519 \citep{Smith+1994}, but they are too cool to produce He\,{\sc ii} emission that would be detectable with SPHEREx.
Potential stellar He\,{\sc ii} emission line sources are O supergiants \citep{Hanson+1996, Morris+1996, BohannanCrowther1999}, albeit with weaker signatures than late-type WN stars \citep[][see their fig.~7]{Lenorzer+2002}. 

The lower panel of Fig.~\ref{cspn_obsuper} presents SPHEREx spectrophotometry of two close relatives of classical WR stars -- He~3--519 (early B Ia$^{+}$) and THA 35-II-153 \citep[O3.5\,If*/WN7, ][]{MaizApellaniz+2016}
--  indeed both have WR catalogue aliases WR31a and WR31-1, respectively. THA~35-II-153 reveals weak He\,{\sc ii} 4.76$\mu$m emission (1.01$\mu$m is not detected - see subpanel). {\it Spitzer}/IRS spectroscopy of He~3--519 is included, revealing strong Pf$\alpha$ emission, with a modest flux offset between SPHEREx and {\it Spitzer}/IRS spectroscopy, likely arising from the associated ejecta nebula \citep{Smith+1994}.

\section{SPHEREx observations of WR candidates}\label{new}

Table~\ref{WRcat} provides an overview of WR candidates discussed in this section, including {\it Gaia} DR3 identifications, aliases from \citet{Mulato+2025}, optical/IR photometry, and inferred absolute magnitudes in the J-band using distances
from \citet{bailer-jones2021} inferred from {\it Gaia} parallaxes. The J-band is favoured over the K$_{s}$-band  in this instance since the K$_{s}$-band can be strongly
affected by hot dust in the case of WC or [WC]-type stars \citep{Crowther+2006CSPN, Williams2019} despite the higher uncertainties in J-band extinctions. Adopted Bayesian priors affect parallax-inferred distances, as discussed by
\citet{RateCrowther2020}, but we adhere to those from \citet{bailer-jones2021} since our targets likely include a mixture of evolved massive and lower mass stars.
Extinctions for non-dusty WR stars, $A_{\rm J}$, are estimated using intrinsic JHK colours from \citet{RossloweCrowther2015} and $A_{\rm J}/A_{\rm K} = 2.5$ \citep{Indebetouw+2005}, and are presented in Table~\ref{WRcat}. In the case of dusty WC stars, since JHK$_{s}$ colours are contaminated by hot dust
-- identified from 2MASS plus {\it Spitzer}/GLIMPSE \citep{Benjamin+2003} or {\it WISE} \citep{WISE} mid-IR photometry, such that {\it Gaia} $G_{\rm BP} - G_{\rm RP}$ colours are preferred to estimate 
$A_{\rm J}$ following \citet[][their figure 9]{RateCrowther2020}. For WR70-19 a parallax-based distance is not known, so this is estimated from weak-lined WN5-6 absolute J-band magnitudes \citep{RateCrowther2020}.

\begin{landscape}
\begin{table}
    \caption{Newly confirmed Galactic Wolf-Rayet stars and related emission line sources from SPHEREx spectrophotometry obtained from {\it Gaia} ESP-ELS candidates \citep{Creevey+2023}, including the numerical value of 
    {\tt classlabel$\_$espels$\_$flag}, ranging from 0 (very high probability) to 4 (low probability). Aliases from \citet[][MMC25]{Mulato+2025} are included. Photometry is from {\it Gaia} ($G_{\rm BP}, G_{\rm RP}$), 2MASS (JHK$_{s}$) and {\it WISE} (W1). Photogeometric distance moduli (DM) are inferred from {\it Gaia} DR3 parallax \citep{bailer-jones2021}, while $A_{\rm J}$ follows \citet{RossloweCrowther2015}, aside from dusty WC stars for which {\it Gaia} observed $G_{\rm BP} - G_{\rm RP}$ colours are used to estimate $A_{\rm J} \sim 2.5 A_{\rm K}$  \citep{RateCrowther2020}. J-band absolute magnitudes, $M_{\rm J}$, are generally inferred from J, $A_{\rm J}$ and DM ($\rightarrow$) with the exception of WR70-19 for which
    DM is inferred from J, $A_{\rm J}$, and an
    adopted $M_{\rm J}$ ($\leftarrow$) from \citet{RateCrowther2020}.}
    \centering
    \label{WRcat}
    \begin{tabular}{
l@{\hspace{2mm}}
l@{\hspace{2mm}}
l@{\hspace{2mm}}
c@{\hspace{2mm}}
c@{\hspace{2mm}}
c@{\hspace{2mm}}
r@{\hspace{2mm}}
r@{\hspace{2mm}}
r@{\hspace{2mm}}
r@{\hspace{2mm}}
r@{\hspace{2mm}}
r@{\hspace{2mm}}
r@{\hspace{2mm}}
r@{\hspace{2mm}}
r@{\hspace{3mm}}
l@{\hspace{2mm}}
c@{\hspace{2mm}}
c@{\hspace{2mm}}
l@{\hspace{2mm}}
l}
    \hline
    ID & ESP- & {\it Gaia} DR3 & MMC25 & RA & Dec & $l$ & $b$ & $G_{\rm BP}$ & $G_{\rm RP}$ & J & H & K$_{s}$ & W1 & DM & $A_{\rm J}$ & & $M_{\rm J}$ & SpT & Notes\\
        & ELS  &        &       & \multicolumn{2}{c}{--- J2000 --- } & deg & deg & mag & mag & mag & mag & mag & mag & mag & mag &  & mag &  & \\
    \hline
    \multicolumn{20}{c}{New Wolf-Rayet stars} \\
    \hline
    WR44-2   & 0  & 5333920438969809152 & WR-C-08 & 11:24:36.53 & --62:32:35.0 & 293.1101 & --1.3388 & 19.42 & 15.27 & 11.95 & 10.71 & 9.85  & 8.94 & 13.1$^{+0.7}_{-0.6}$& 2.0$^{+0.2}_{-0.2}$ & $\rightarrow$ & --3.1$^{+0.7}_{-0.6}$ & WN4-6b & $\cdots$ \\ [2pt] 
     WR62-3 & 0  & 5878372204454901248 & WR-C-15 & 14:35:08.70 & --61:14:02.5 & 315.0793 & --0.8323 & 19.32 & 15.45 & 12.68 & 11.81 & 11.25 & 10.58 & 13.8$^{+0.6}_{-0.6}$ & 2.0$^{+0.2}_{-0.2}$ & $\rightarrow$ & --3.1$^{+0.6}_{-0.6}$ & WN3 & $\cdots$ \\ [2pt] 
    WR70-18  & 0 & 5882603507496595200 & WR-C-19 & 15:37:41.30 & --57:23:35.8 & 323.8861 & --1.4482 & 17.99 & 14.94 & 12.80 & 11.99 & 11.18 & 10.44 & 14.4$^{+0.6}_{-0.4}$ & 1.5$^{+0.2}_{-0.2}$ & $\rightarrow$ & --3.1$^{+0.6}_{-0.4}$  & WC7--8 & $\cdots$ \\ [2pt] 
     WR70-19 & 1 & 5884578986303334016 & $\cdots$ & 15:55:08.61 & --54:59:20.7 & 327.2791 & --1.0483 & 16.97 & 14.25 & 12.01: & 11.27 & 10.74 & 9.57 & 13.8$^{+0.7}_{-0.7}$ & 2.0$^{+0.7}_{-0.7}$ & $\leftarrow$ & --3.8$^{+0.7}_{-0.7}$ & WN6--7 & $\cdots$ \\ [2pt] 
    WR77-9  & 0 & 5939996017261315968 & WR-C-22 & 16:49:41.87 & --46:45:08.7 & 339.1529 & --1.3322 & 19.76 & 14.71 & 11.27 & 10.06 & 9.13 & 8.12 & 13.3$^{+0.7}_{-0.6}$ & 3.0$^{+0.3}_{-0.3}$ & $\rightarrow$ & --$5.0^{+0.7}_{-0.6}$ & WC9d & $\cdots$\\ [2pt] 
    WR84-13 & 1  & 5971904330062036224 & WR-C-25 & 17:14:00.39 & --40:30:42.1 & 346.7762 & -0.9822 & 19.80 & 15.72 & 12.60 & 11.39 & 10.33 & 9.13 & 13.1$^{+1.0}_{-0.6}$ & 2.5$^{+0.2}_{-0.2}$ & $\rightarrow$ & --3.0$^{+1.0}_{-0.6}$ & WC9d & $\cdots$ \\ [2pt] 
    WR84-14 & 0 & 5965760637051514496 &  $\cdots$ & 17:14:24.11 & --41:57:25.1 & 345.6467 & --1.8873 & 17.45 & 14.70 & 12.27: & 11.33 & 10.81 & 10.24 & 13.8$^{+0.3}_{-0.6}$ & 1.5$^{+0.2}_{-0.2}$ & $\rightarrow$ & --3.0$^{+0.4}_{-0.6}$ & WN7 & $\cdots$ \\ [2pt] 
    WR94-5 & 1 & 4054402569982187904 & WR-C-27 & 17:33:23.11 & --33:43:52.5 & 354.5509 & --0.3424 & 18.20 & 14.68 & 12.02 & 11.23 & 10.66 & 10.01 & 13.6$^{+0.4}_{-0.3}$ & 1.4$^{+0.2}_{-0.2}$ & $\rightarrow$ & --3.0$^{+0.5}_{-0.4}$ & WN6 & $\cdots$ \\ [2pt] 
    WR98-3 & 2  & 4056935402389532672 & WR-C-30 & 17:41:27.51 & --30:16:42.6 & 358.3827 & +0.0654 & 20.50 & 14.83 & 10.27 & 7.93 & 6.10 & 3.61 & 13.5$^{+0.6}_{-0.7}$ & 3.2$^{+0.3}_{-0.3}$ & $\rightarrow$ & --6.4$^{+0.6}_{-0.7}$ & WC9d & $\cdots$ \\ [2pt] 
    WR111-16  & 0 & 4066281766676330368 & WR-C-33 & 18:08:52.12 & --23:55:27.1 & 6.9458 & --1.9871 & 19.02 & 14.70 & 11.56 & 10.24 & 9.04 & 7.91 & 13.6$^{+0.6}_{-0.6}$ & 2.8$^{+0.3}_{-0.3}$ & $\rightarrow$ & --4.8$^{+0.6}_{-0.6}$ & WC4--7d & $\cdots$ \\ [2pt] 
     WR125-7 & 1 & 2021131933611057536 & WR-C-51 & 19:38:34.47 & +24:03:20.5 & 59.5413 & +1.1401 & 19.76 & 16.23 & 13.56 & 12.66 & 11.99 & 11.02 & 14.3$^{+0.5}_{-0.5}$ & 1.7$^{+0.2}_{-0.2}$ & $\rightarrow$ & --2.5$^{+0.5}_{-0.5}$ & WN5--6 & $\cdots$ \\ [2pt] 
    WR128-1  & 1 & 1828324426391283200 & WR-C-52 & 19:52:40.81 & +23:29:33.9 & 60.6724 & --1.9343 & 18.03 & 15.26 & 13.32 & 12.56 & 12.01 & 11.05 & 14.1$^{+0.4}_{-0.7}$ & 1.5$^{+0.2}_{-0.2}$ & $\rightarrow$ & --2.3$^{+0.4}_{-0.7}$ & WN4--5 & $\cdots$ \\ [2pt] 
    WR148-3  & 1 & 2169491477039398144 & WR-C-56 & 21:05:28.26 & +51:36:00.8 & 91.6922 & +2.9904 & 19.73 & 15.82 & 12.91 & 11.84 & 11.06 & 10.19 & 14.2$^{+0.5}_{-0.9}$ & 1.8$^{+0.2}_{-0.2}$ & $\rightarrow$ & --3.1$^{+0.5}_{-0.9}$ & WN4b & $\cdots$ \\ [2pt] 
    \hline
     \multicolumn{20}{c}{New Wolf-Rayet CSPNe} \\
    \hline
   G307.2-0.6 & 0 & 5865253866850321280 & WR-C-13 & 13:29:11.16 & --63:10:09.2 & 307.1763 & --0.6090 & 19.75 & 15.64 & 12.72 & 11.67 & 10.74: & 10.01 & 12.7$^{+0.3}_{-0.3}$ & 1.9$^{+0.2}_{-0.2}$ & $\rightarrow$ & --1.9$^{+0.4}_{-0.4}$ & [WC8--9] & $\cdots$ \\ [2pt] 
   G326.5-0.6 &  1 & 5884420446123080320 & WR-C-20 & 15:48:50.49 & --55:11:20.2 & 326.4534 & --0.6345 & 18.98 & 14.31 & 10.54 & 8.52 & 6.94 & 5.57 & 8.9$^{+0.2}_{-0.2}$ & 2.7$^{+0.3}_{-0.3}$ & $\rightarrow$ & --1.1$^{+0.4}_{-0.4}$ & [WC9] & $\cdots$ \\ [2pt] 
   G31.2-1.3 & 0 & 4258982586850246784 & WR-C-45 & 18:53:05.40 & --02:08:33.2 & 31.2130 & --1.3442 & 19.55 & 15.78& 13.29 & 12.45 & 11.56 & 10.97 & 12.7$^{+0.5}_{-0.7}$ & 1.7$^{+0.2}_{-0.2}$ & $\rightarrow$ & --1.1$^{+0.5}_{-0.7}$ & [WC7--8] & $\cdots$ \\ [2pt] 
    \hline
     \multicolumn{20}{c}{Other emission-line sources} \\
    \hline
    PMR 1 & 4 & 5313911079687293440 & $\cdots$ & 09:28:40.97  & --49:36:46.6 & 272.8387 & +1.0325 & 17.75 & 16.03 & 14.19 & 13.56 & 13.15  & 12.38 & 11.8$^{+0.4}_{-0.4}$ & 0.8$^{+0.2}_{-0.2}$ & $\rightarrow$ & +1.6$^{+0.4}_{-0.4}$ & [WC] &  $\cdots$ \\ [2pt] 
  $\cdots$ & 2 & 5258760366715078656 & WR-C-06 & 10:15:21.03 & --57:07:05.9 & 282.9434 & --0.4324 & 16.83 & 14.00 & 11.67 & 10.32 & 8.83 & 6.91 & 13.5$^{+0.2}_{-0.2}$ & 4.1$^{+0.8}_{-0.8}$ & $\rightarrow$ & --5.9$^{+0.8}_{-0.8}$ & $\cdots$ & He\,{\sc i} em. \\ [2pt] 
    WR85a & 0 & 5973474703218558080 & $\cdots$ & 17:16:02.18 & --38:15:42.7 & 348.8328 & +0.0110 & 13.88 & 11.60 & 9.71 & 8.43 & 6.83 & 4.84 & 12.5$^{+0.1}_{-0.1}$ & 4.3$^{+1.0}_{-1.0}$ & $\rightarrow$ & --7.1$^{+1.0}_{-1.0}$ & $\cdots$ & He\,{\sc i} em. \\  [2pt] 
    $\cdots$ & 3 & 4095106941425872640 & WR-C-34 & 18:09:48.33 & --19:18:00.7 & 11.0988 & +0.0594 & 18.49 & 15.35 & 12.47 & 10.80 & 9.38 & 7.61 & 12.4$^{+0.3}_{-0.2}$ & 3.8$^{+0.3}_{-0.3}$ & $\rightarrow$ & --3.7$^{+0.4}_{-0.4}$ & WC? & C\,{\sc iii} em.\\ [2pt] 
    J1823+0250 & 2 & 4277897038692284032 & $\cdots$ & 18:23:47.04 & +02 50 17.5 & 32.3107 & +7.4450 & 16.17 & 15.19 & 14.60 & 14.41 & 14.33 & $\cdots$ & 13.9$^{+0.3}_{-0.4}$ & 0.8$^{+0.2}_{-0.2}$ & $\rightarrow$ 
   & --0.1$^{+0.4}_{-0.4}$ & WN-like  &  He\,{\sc ii} em \\ [2pt]  
    NGC~6751 & 3 & 4206136209740612352 & $\cdots$ & 19:05:55.54 & --05:59:32.9 & 29.2275 & --5.9418 & 13.61 & 13.81 & 13.37 & 13.23 & 12.63 
    & 10.02 & 12.5$^{+0.3}_{-0.3}$ & 0.1$^{+0.1}_{-0.1}$ & $\rightarrow$ & +0.8$^{+0.4}_{-0.4}$ & [WC] & $\cdots$ \\ [2pt]    
J1907-0523 & 3 & 4206279695997383040 & $\cdots$ & 19:07:39.60 & --05:23:33.3 & 29.9613 & --6.0570 & 16.74 & 16.07 & 15.33 & 15.26 & 15.27 & 14.55 & 14.3$^{+0.4}_{-0.4}$ & 0.6$^{+0.1}_{-0.1}$ & $\rightarrow$ & +0.4$^{+0.4}_{-0.4}$ 
    & WN-like &  He\,{\sc ii} em \\ [2pt] 
    $\cdots$ & 2 & 2059241178412827520 & WR-C-53 & 20:09:08.72 & +36:38:00.9 & 73.7111 & +1.9795 & 17.74 & 15.66 &  14.50 & 13.92 & 13.39 & 11.66 & 14.2$^{+0.3}_{-0.5}$ & 1.7$^{+0.2}_{-0.2}$ & $\rightarrow$ & --1.4$^{+0.4}_{-0.5}$ & WN? & He\,{\sc ii} em? \\ [2pt] 
    \hline
    \end{tabular}
    \end{table}
\end{landscape}



\subsection{High probability candidates}

We have extracted SPHEREx spectrophotometry of Wolf-Rayet candidates with {\tt classlabel$\_$espels$\_$flag} in the range 0 to 2, excluding: (i) known WR stars; 
(ii) other stellar types (e.g. Young Stellar Objects, Cataclysmic Variables);  (iii) candidates spectroscopically observed by \citet{Mulato+2025}; (iv)
sources at  high galactic latitude i.e. $|b|\geq 10^{\circ}$; (v) peculiar emission line sources (e.g. HD 45166, WR85a). HD~45166 has historically been described as a quasi-WR star, but \citet{Shenar+2023} firmly established it as a highly magnetic He star, resulting from a merger. Low resolution optical spectroscopy of WR85a closely resembles the unusual He-rich, nebular dominated source WR122 \citep{NaSt1, Mauerhan+2015}. WR85a is known to exhibit  He\,{\sc i} 1.08$\mu$m emission \citep{Morris+1996}, although its nature remains uncertain. 

\begin{figure*}
    \centering
     \includegraphics[width=0.9\columnwidth,angle=0]{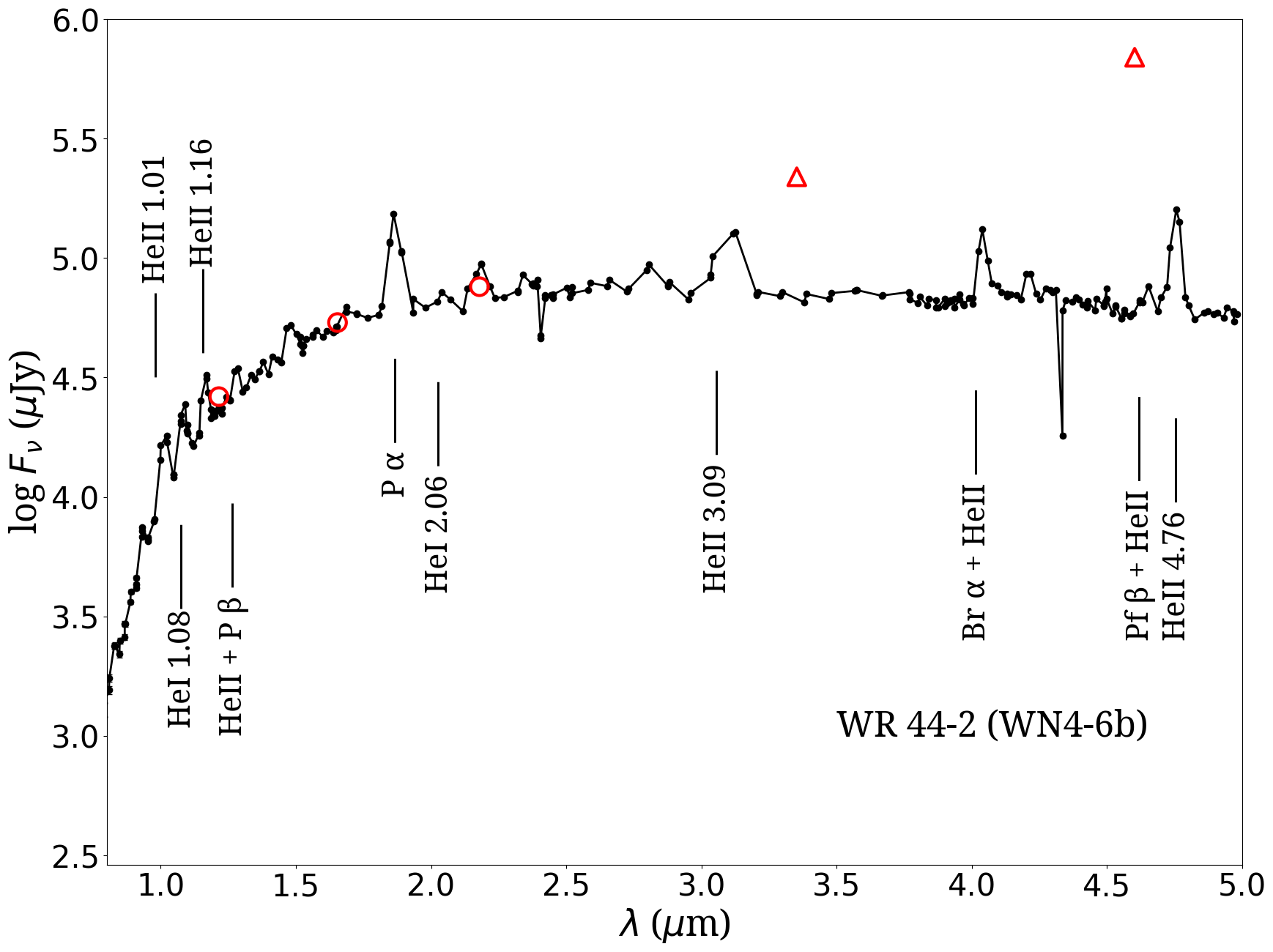}
     \includegraphics[width=0.9\columnwidth,angle=0]{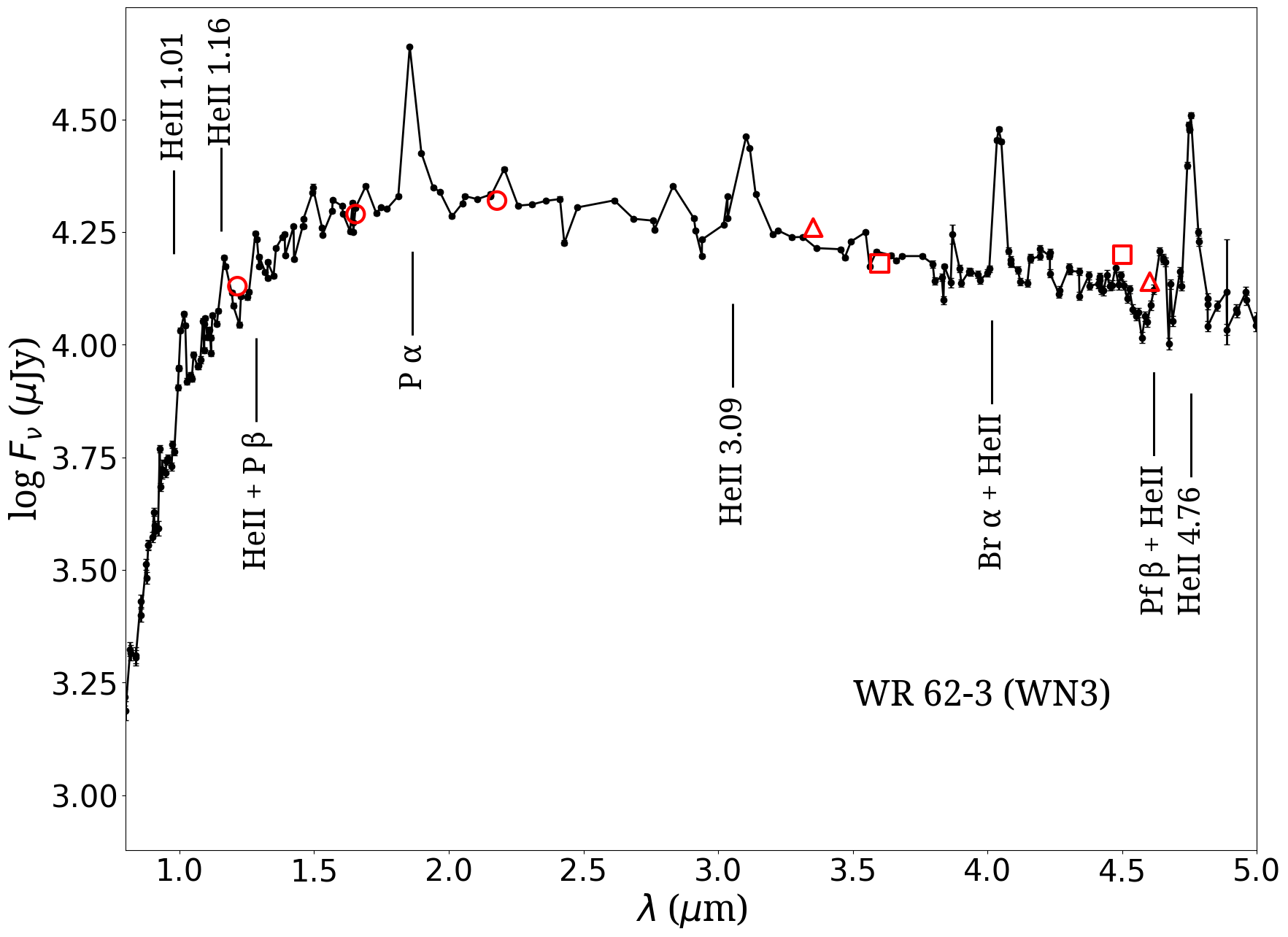}
     \includegraphics[width=0.9\columnwidth,angle=0]{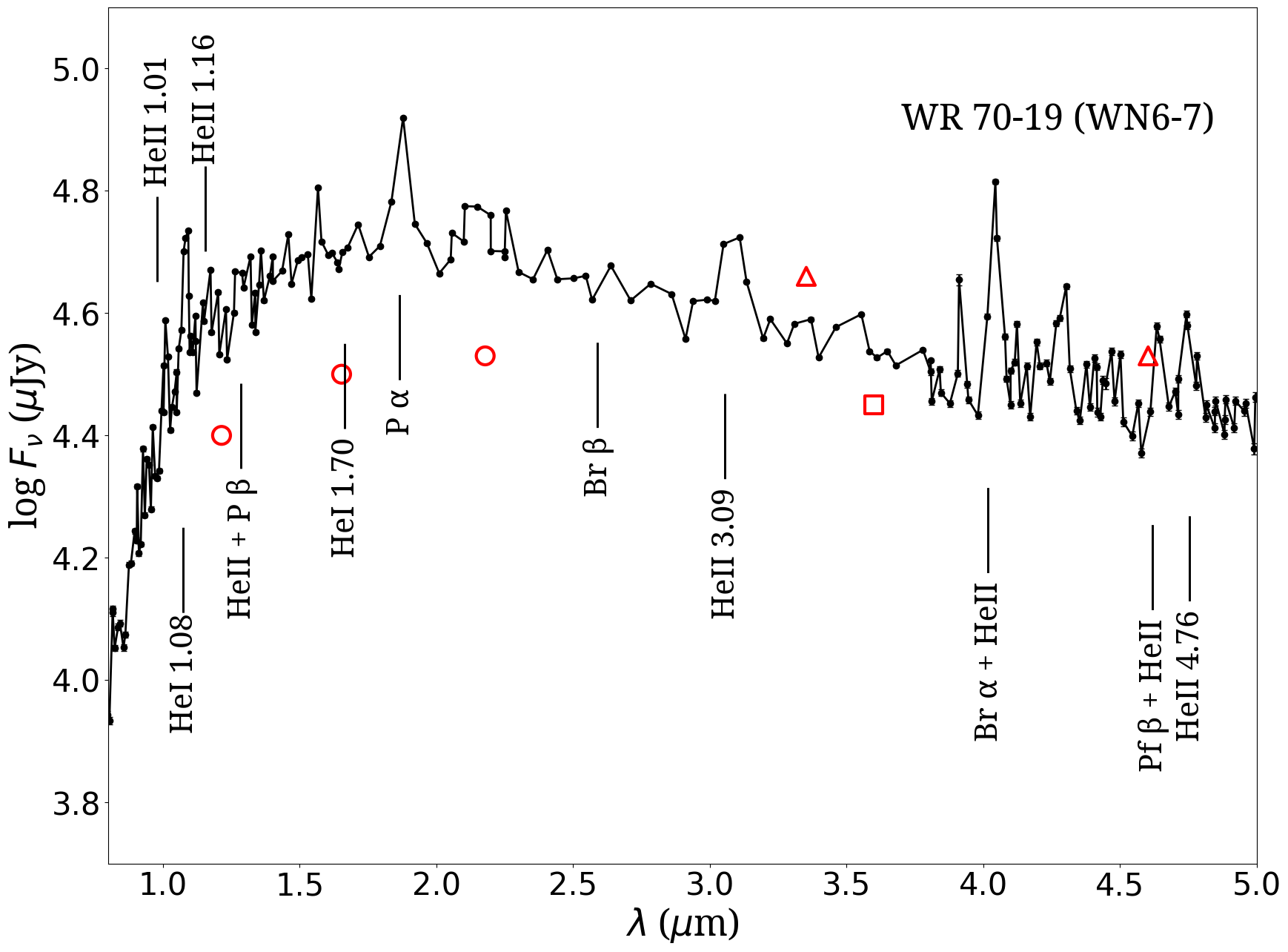}
     \includegraphics[width=0.9\columnwidth,angle=0]{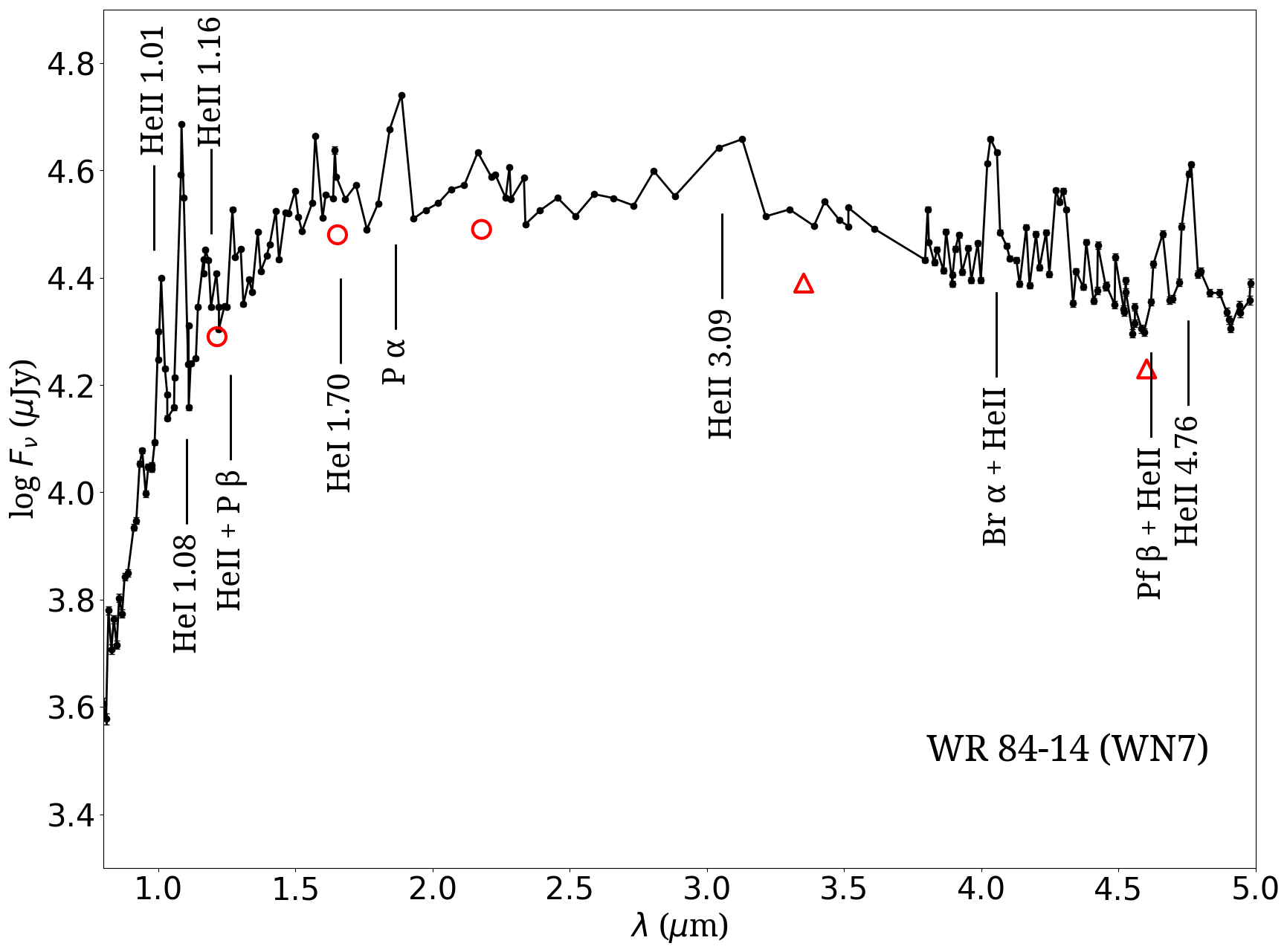}
     \includegraphics[width=0.9\columnwidth,angle=0]{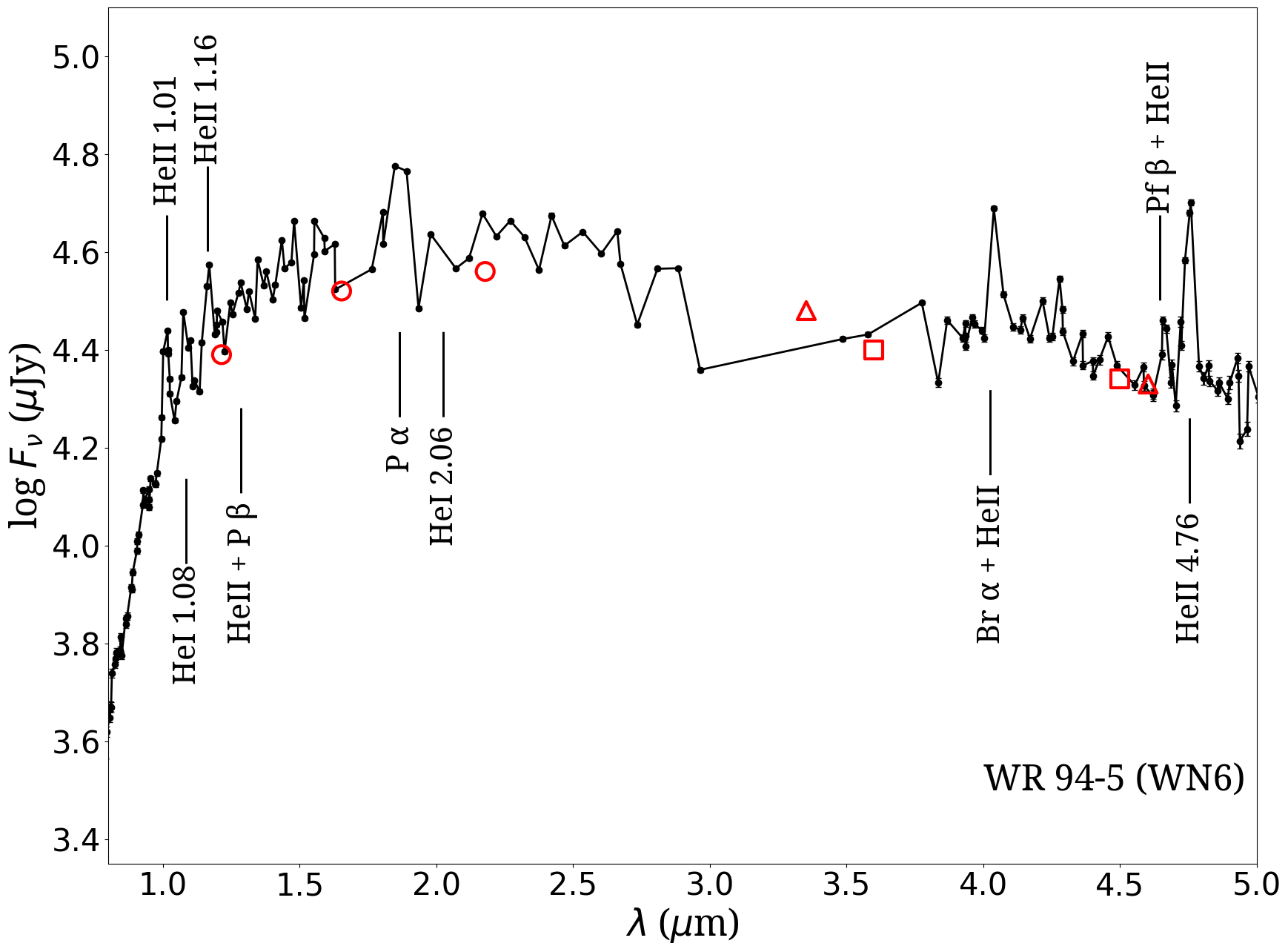}
     \includegraphics[width=0.9\columnwidth,angle=0]{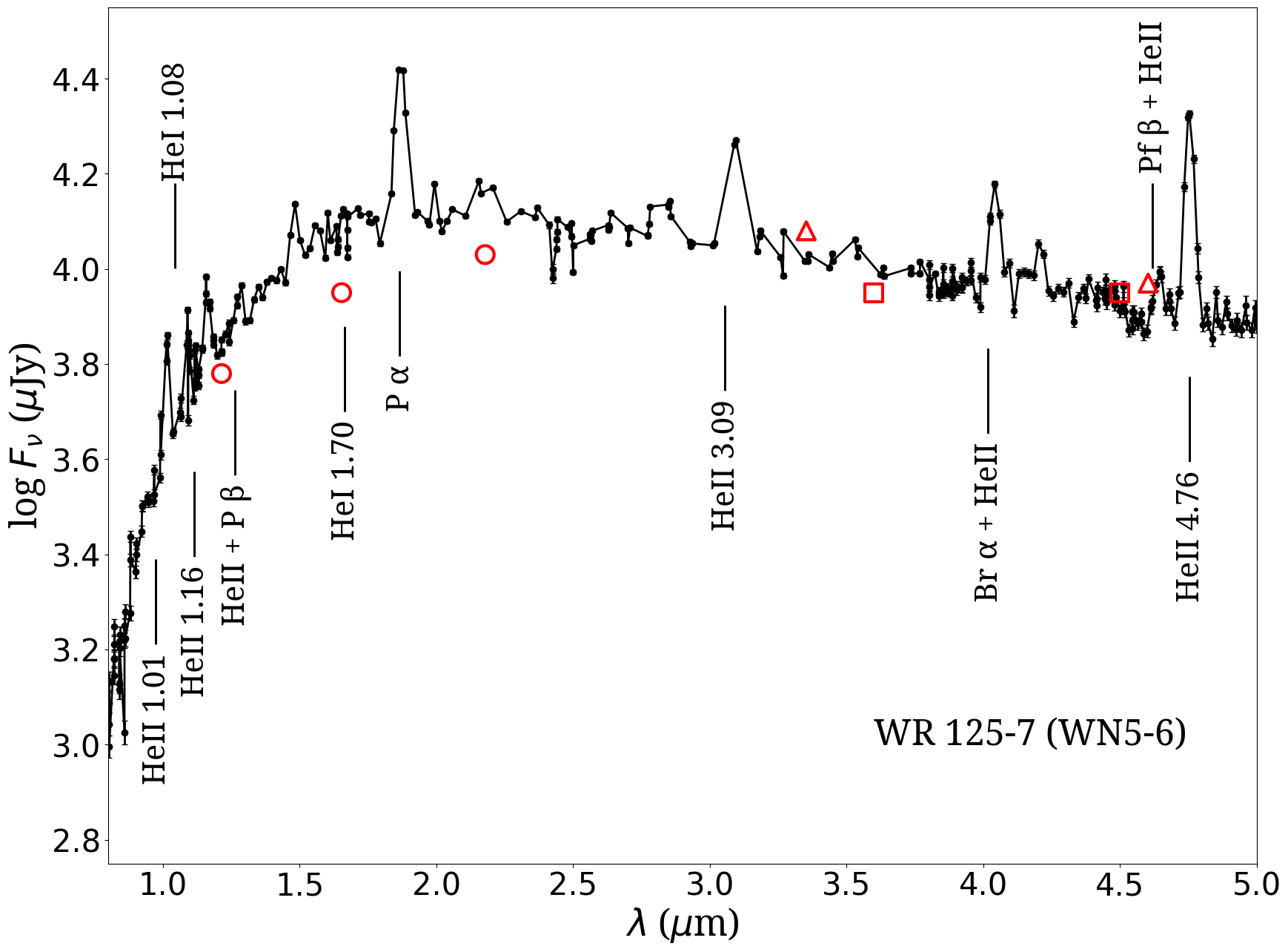}
     \includegraphics[width=0.9\columnwidth,angle=0]{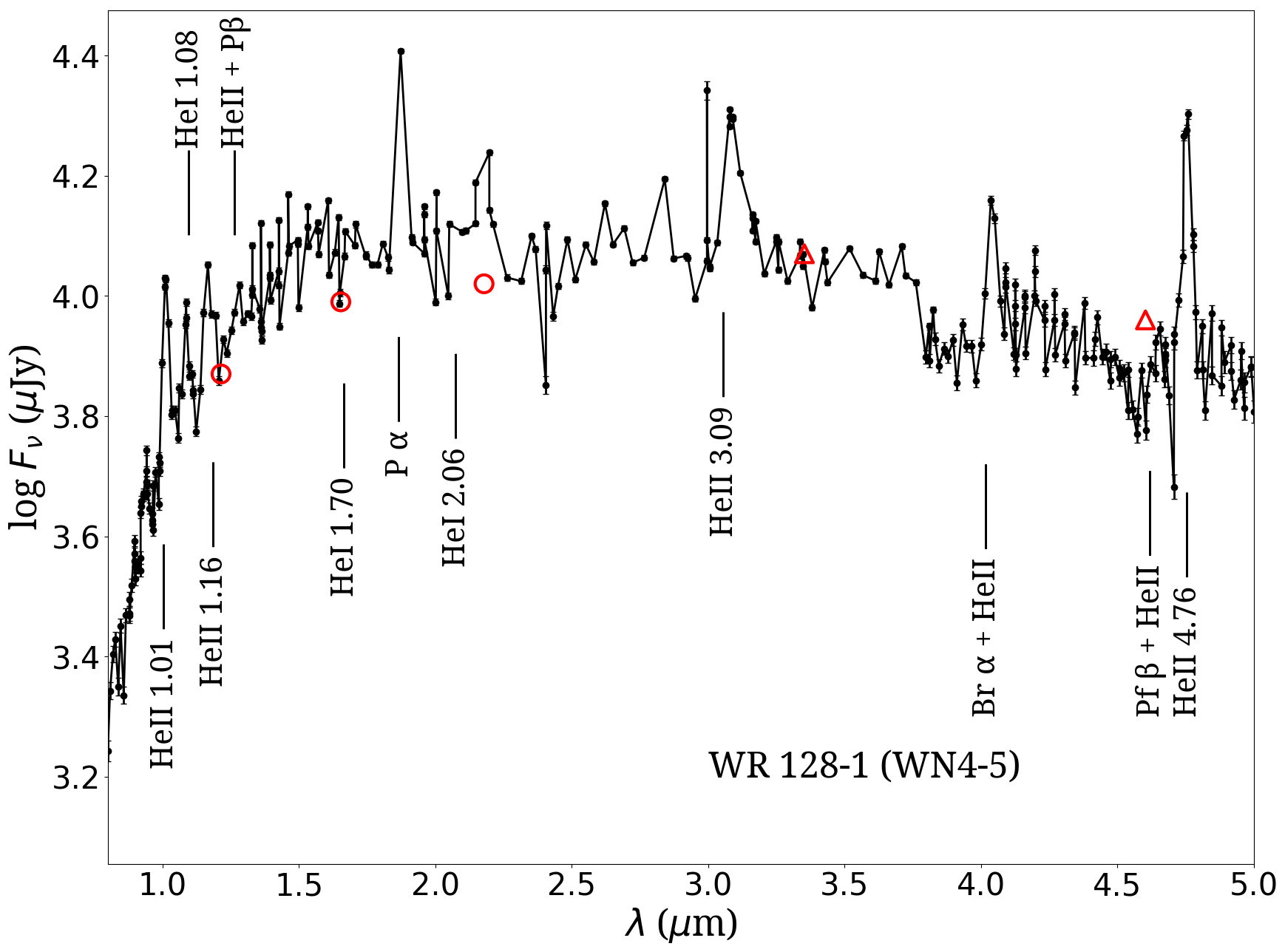}
     \includegraphics[width=0.9\columnwidth,angle=0]{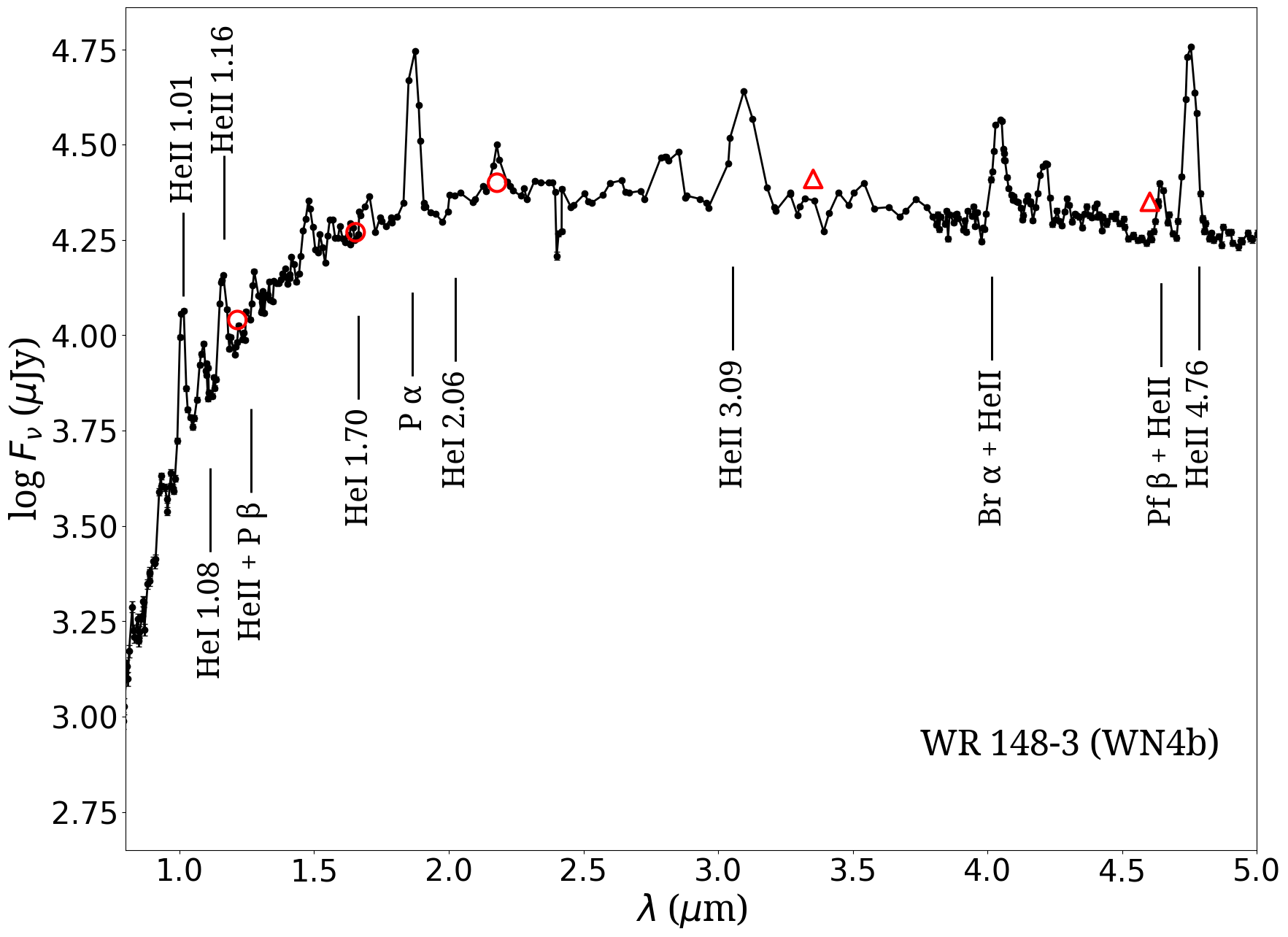}
         \caption{SPHEREx spectrophotometry of newly identified WN stars, together with photometry from 2MASS (open circles), {\it Spitzer}/GLIMPSE (open squares) and {\it WISE} (open triangles). Fluxes are in logarithmic units given the wide dynamic range of spectrophotometry.}
         \label{WN-stars}
    \end{figure*}

\begin{figure*}
    \centering
      \includegraphics[width=0.9\columnwidth,angle=0]{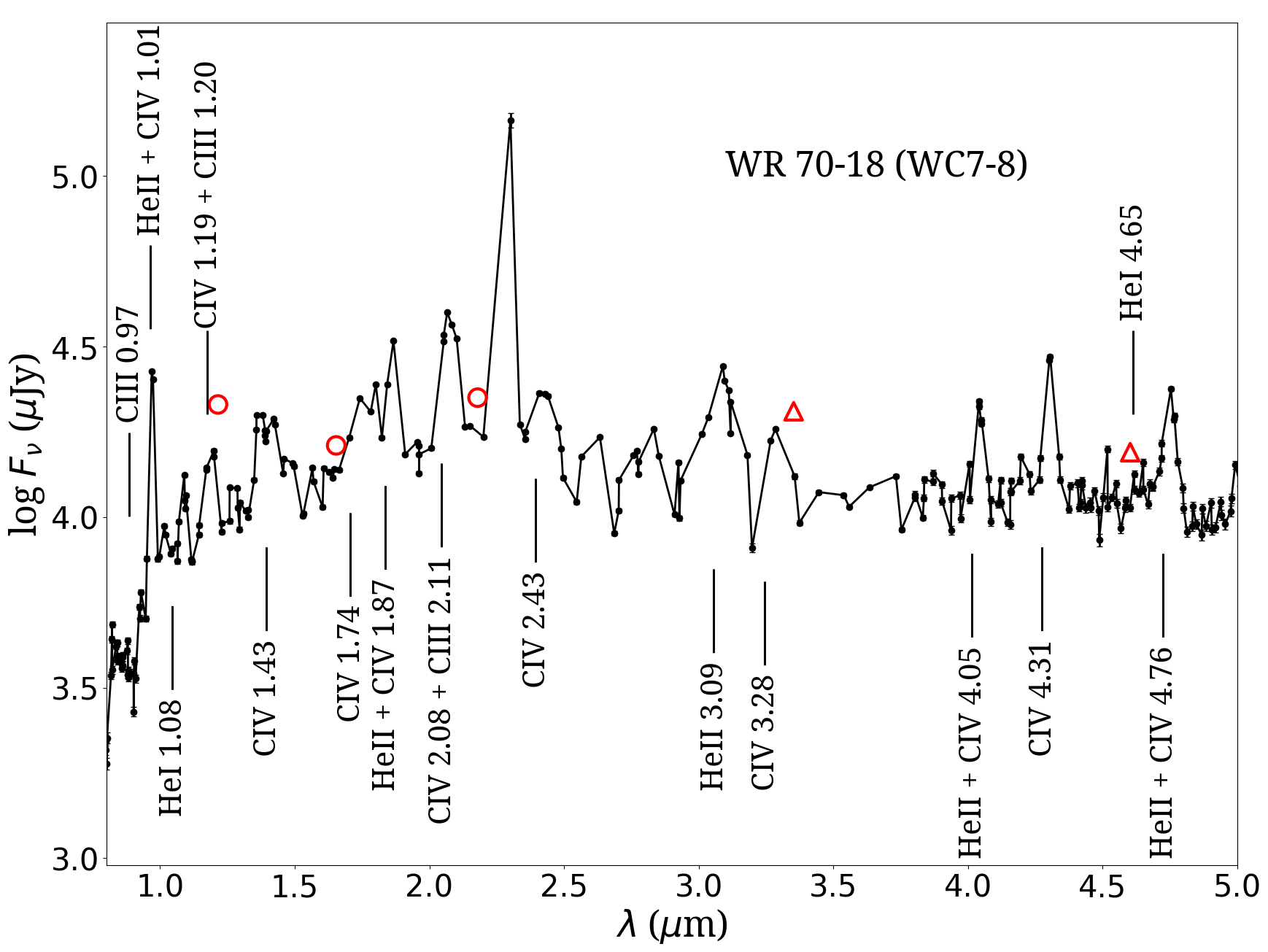}
       \includegraphics[width=0.9\columnwidth,angle=0]{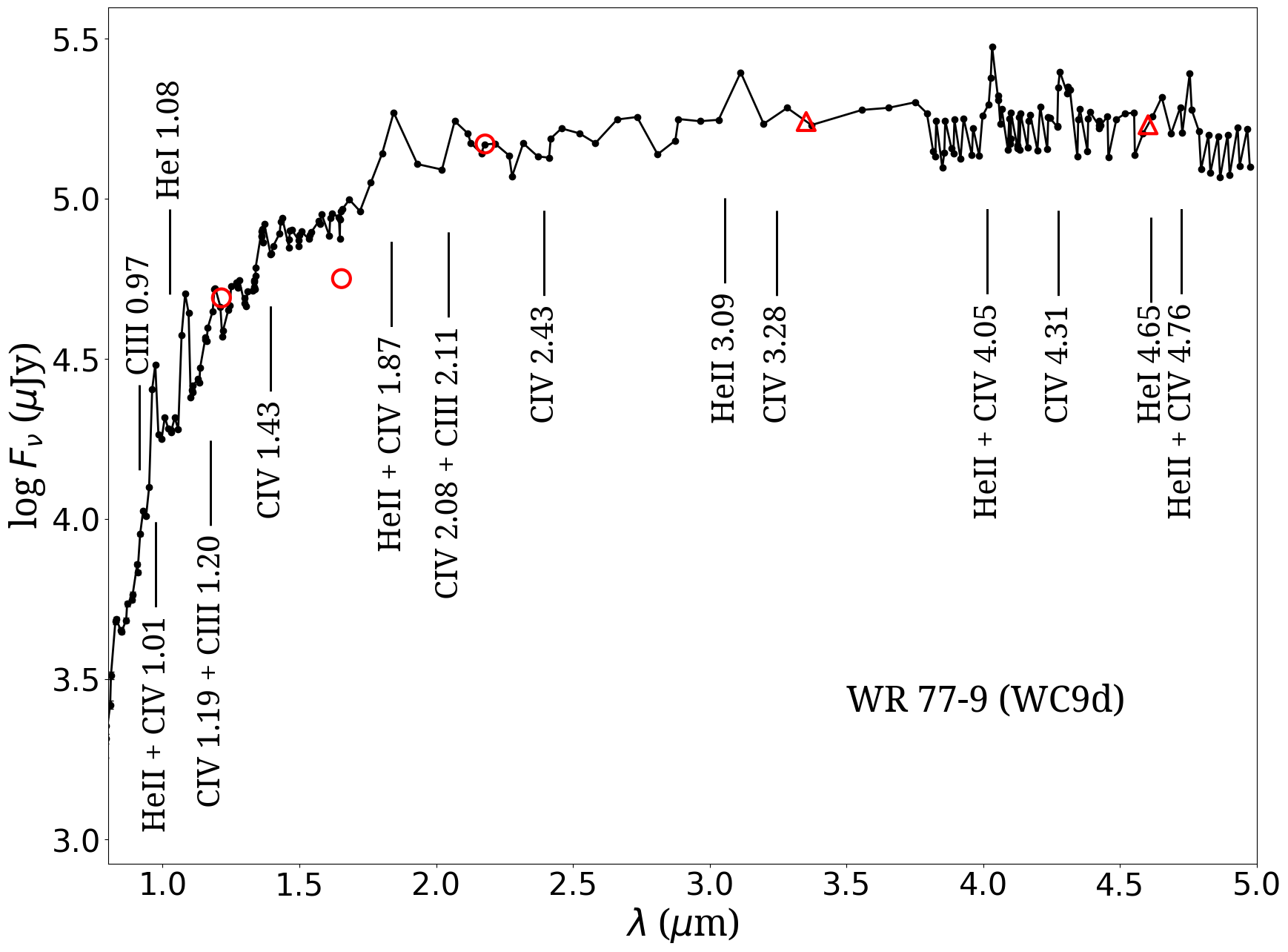}
     \includegraphics[width=0.9\columnwidth,angle=0]{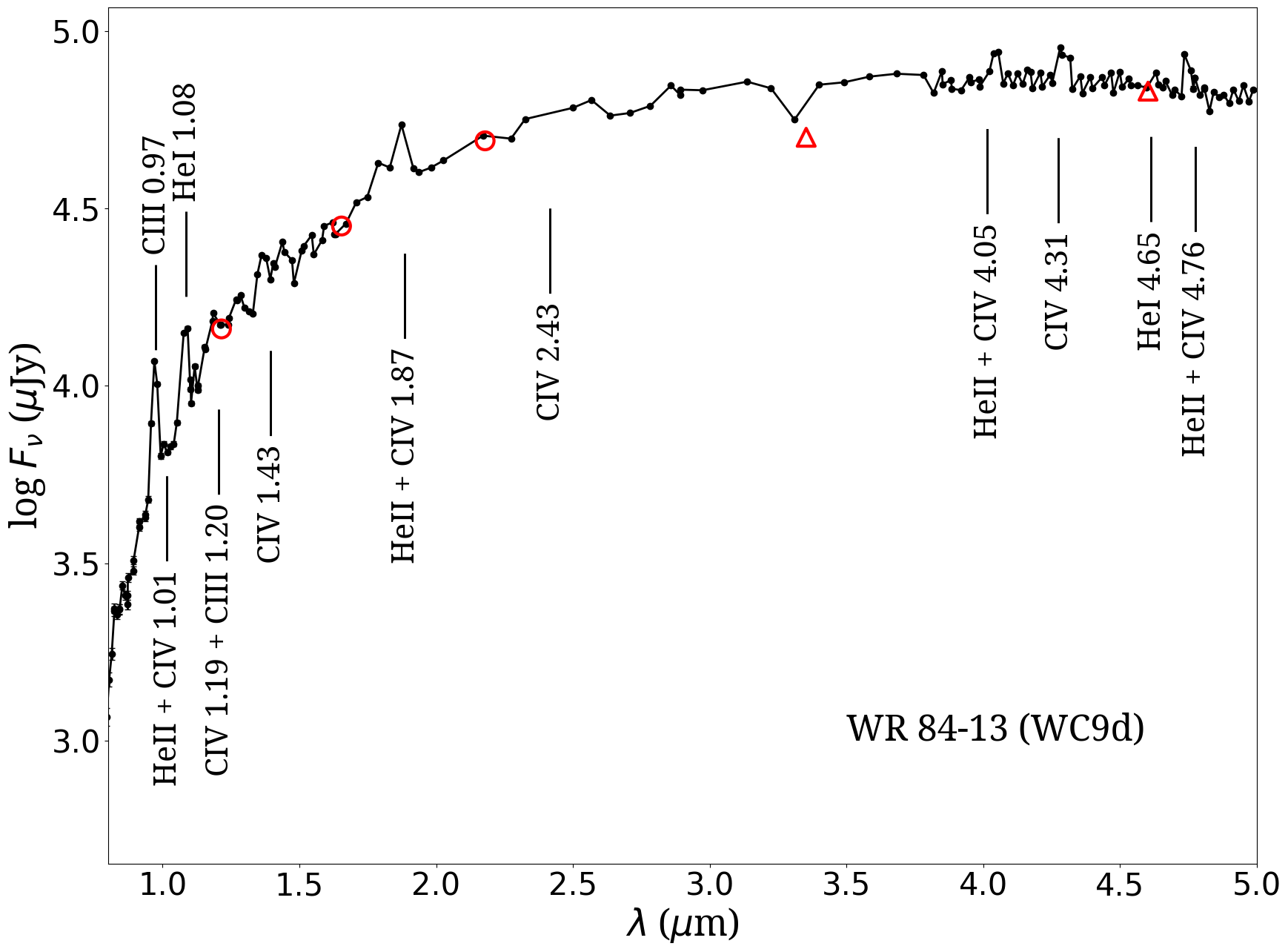}
     \includegraphics[width=0.9\columnwidth,angle=0]{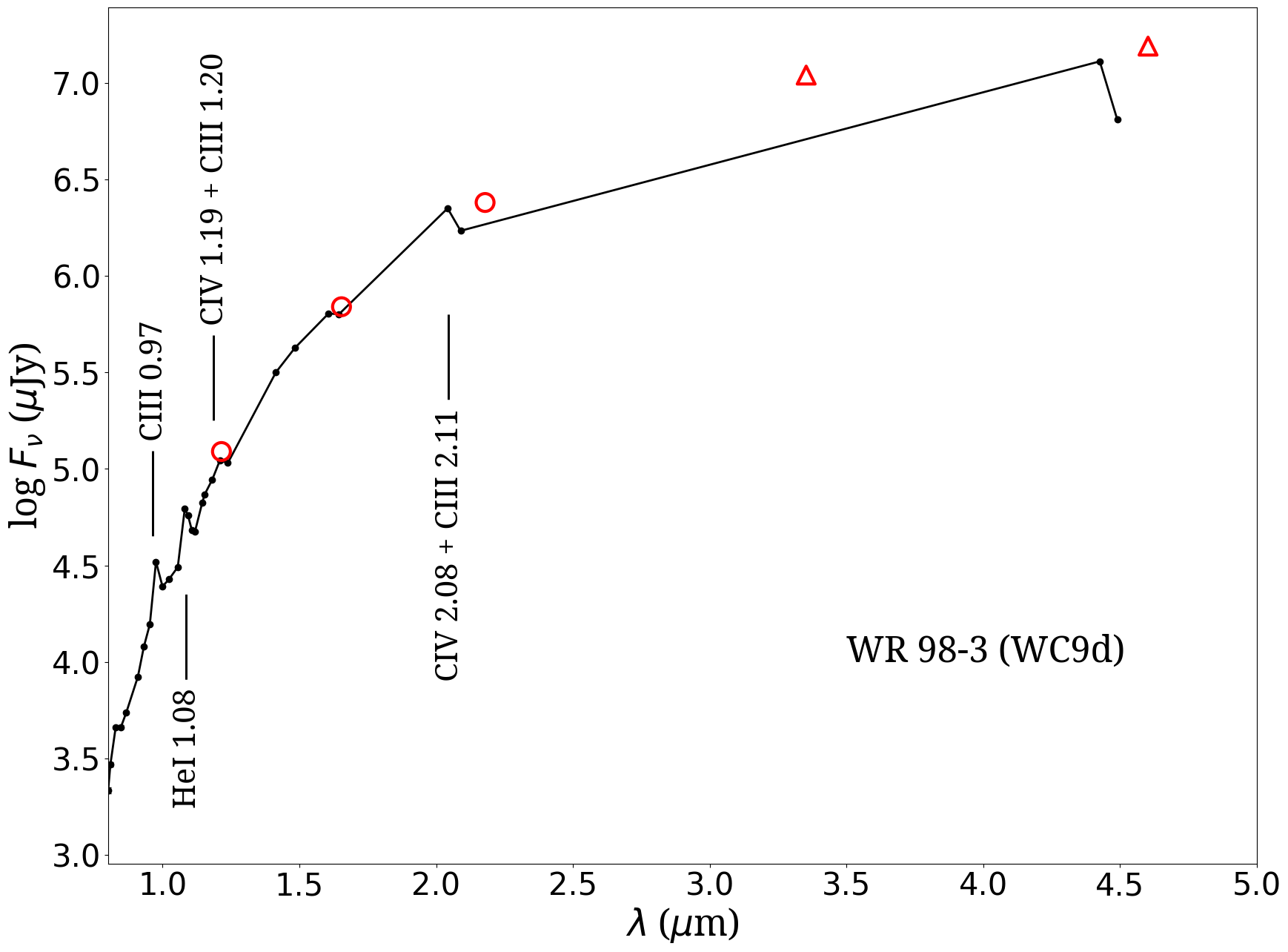}
          \includegraphics[width=0.9\columnwidth,angle=0]{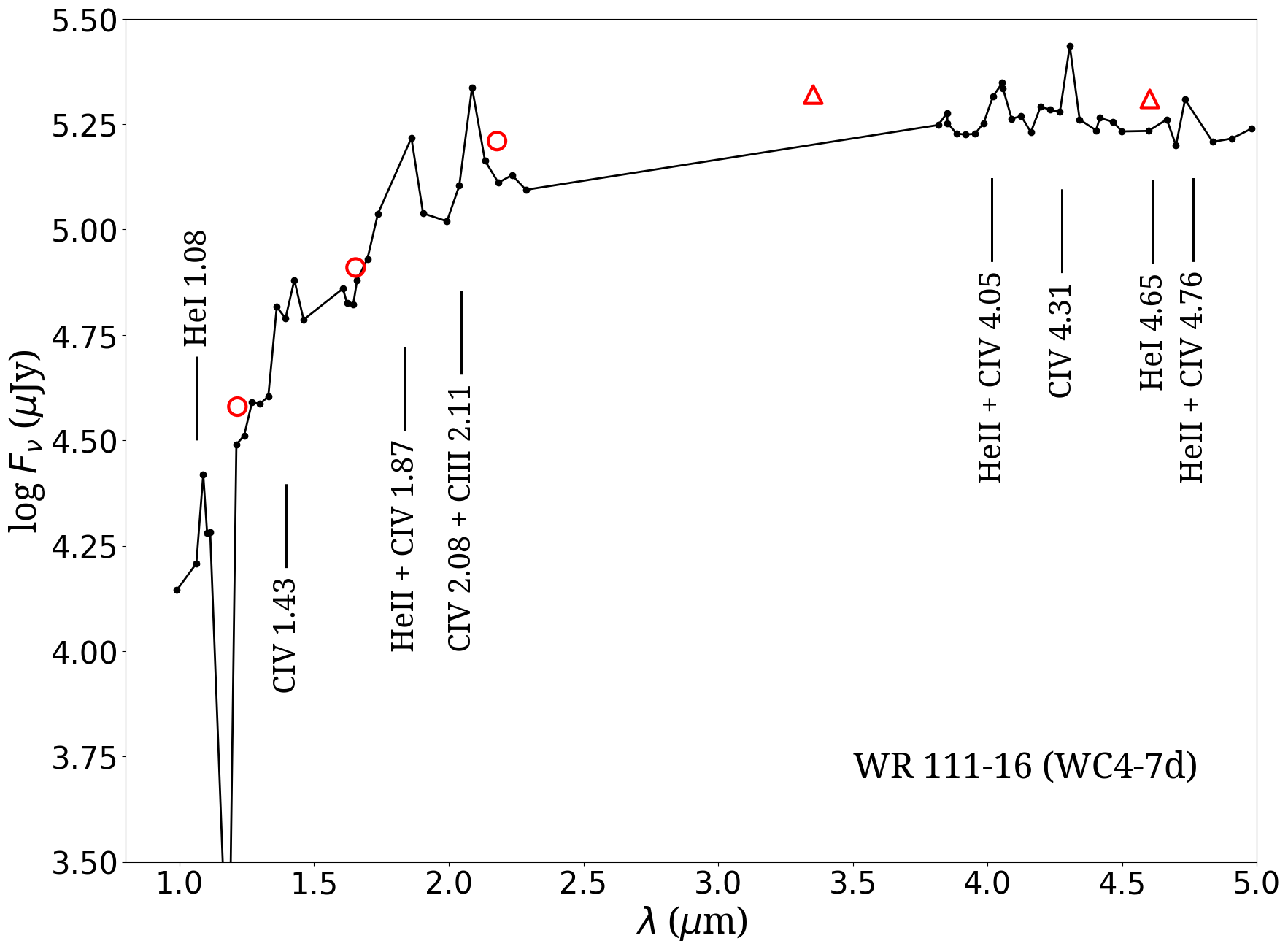}
       \caption{SPHEREx spectrophotometry of newly identified WC stars, together with photometry from 2MASS (open circles) and {\it WISE} (open triangles). Fluxes are in logarithmic units given the wide dynamic range of spectrophotometry.}
         \label{WC-stars}
    \end{figure*}

13 of the 22 remaining candidates are confirmed as Wolf-Rayet stars -- 8 WN and 5 WC stars -- which are presented in Fig.~\ref{WN-stars} and Fig.~\ref{WC-stars}, respectively. Photometry from 2MASS \citep[open circles,][]{skrutskie2006}, {\it Spitzer}/GLIMPSE \citep[open squares,][]{Benjamin+2003}, and {\it WISE}  \citep[open triangles,][]{WISE} is included. SPHEREx spectrophotometry generally matches photometry, with the exception of {\it WISE} photometry of WR44-2. Assigned WR catalogue numbers follow the 2012 IAU Working Group for Massive Star recommendations presented in Appendix A of \citet{RossloweCrowther2015}, i.e. WR44-2 denotes the second identified Galactic WR whose Right Ascension lies between WR44 and WR45 \citep{vanderHucht2001}.

The low resolution SPHEREx spectrophotometry hinders IR spectral classification \citep{RossloweCrowther2018}, due to line blends (e.g. He\,{\sc i} 1.70$\mu$m/He\,{\sc ii} 1.69$\mu$m, C\,{\sc iii} 1.20$
\mu$m/C\,{\sc iv} 1.19$\mu$m) and distinguishing narrow-lined from broad-lined WN stars. Nevertheless, estimates of the emission equivalent width of He\,{\sc ii} 1.01$\mu$m permits 
strong-lined early WN stars to be distinguished from weak-lined early WN stars \citep{Smith+1996}, and the He\,{\sc i} 1.08$\mu$m/He\,{\sc ii} 1.01$\mu$m ratio (recall insets in Fig.~\ref{SPHEREx-WR}) permits spectral type estimates \citep{RossloweCrowther2018}. Since emission lines may be diluted by other IR sources in the vicinity of the WR star, the SPHEREx band (B6) with the highest resolving power includes He\,{\sc ii} 4.76$\mu$m. WR136 narrowly qualifies as a strong, broad-lined WN star (WNb) with FWHM(He\,{\sc ii} 4.76$\mu$m) $\sim$ 2500 km\,s$^{-1}$, uncorrected for instrumental broadening, below which WN stars are classified as weak, narrow-lined.

From the spectrophotometry presented in Figure~\ref{WN-stars} we identify WR44-2 and  WR148-3 as strong-lined (WNb) early-type stars, with WR62-3, WR70-20, WR84-14, WR94-5, WR128-1 and WR125-7 as weak-lined 
stars, ranging from WN3 (WR62-3) to WN7 (WR84-14). 

For WC stars, the relative
strengths of C\,{\sc iii} 0.97$\mu$m, He\,{\sc ii} 1.01$\mu$m and He\,{\sc i} 1.08$\mu$m provide the primary diagnostics (recall insets in Fig.~\ref{SPHEREx-WR}), since the ratio of C\,{\sc iv} 2.08$\mu$m to  C\,{\sc iii} 2.11$\mu$m recommended by \citet{RossloweCrowther2018} is severely hindered by the low resolution of SPHEREx band B3. From the spectrophotometry presented in Figure~\ref{WC-stars} we identify WR111-16 as a WC4--7, WR70-18 as a WC7--8 star since C\,{\sc iii} $\gg$ He\,{\sc i} $\gg$ He\,{\sc ii}, plus WR77-9, WR84-13 and WR98-3 as WC9 stars owing to C\,{\sc iii} $\sim$ He\,{\sc i} $\gg$ He\,{\sc ii}. Rising flux levels from K$_{s}$-band to GLIMPSE [3.6] or {\it WISE} W1 3.5$\mu$m suggest the presence of hot dust (WC $\rightarrow$ WCd), with WR98-3 the most obvious example in Fig.~\ref{WC-stars}.

Turning to the remaining ESP-ELS Wolf-Rayet candidates, three systems match those of WC stars albeit with unrealistically faint absolute magnitudes. SPHEREx spectrophotometry of these sources is presented in the Appendix (Fig.~\ref{CSPNe}), together with the Planetary Nebula NGC~2392, an early [WC]-type central star PMR~1 \citep{Morgan+2001} plus a late WC star. WR-C-20 \citep{Mulato+2025} resembles a WC9 star, though has a suspiciously low
distance (0.6 kpc). This has remained undetected to date owing to its very high dust extinction, as evidenced by its very red {\it Gaia} colour 
$G_{\rm BP} - G_{\rm RP}$ = 4.67 mag. Examples of nearby, highly reddened WR systems exist e.g. WR70-16 \citep[Apep,][]{Callingham+2019, Callingham+2020}. Nevertheless, the absolute J-band magnitude of WR-C-20, 
$M_{\rm J}$ = --1.1$\pm$0.4 mag, is over three magnitudes fainter than typical WC9 stars \citep{RateCrowther2020}, so we favour a [WC]-type CSPNe origin,
and subsequently refer to it by its galactic coordinates G326.5--0.6, adding to the catalogue of low-galactic latitude [WC] stars  \citep{Kanarek+2017}.  

A coarse estimate of the bolometric luminosity of G326.5--0.6 can be made from comparison with Potsdam WR (PoWR) model atmosphere
\citep{Grafener+2002, HamannGrafener2003} grids of WC stars \citep{Sander+2012}. The Galactic WC~06-12 model\footnote{https://www.astro.physik.uni-potsdam.de/PoWR/powrgrid1.php}, with properties appropriate to WC9 stars \citep{RateCrowther2020}, has BC$_{J}$ = --3.6 mag, so the estimated luminosity for G326.5--0.6 is $\log L/L_{\odot} \sim$  3.8, typical of CSPNe \citep{Blocker1995, MillerBertolami2016}.  We have also estimated absolute J-band magnitudes of well known CSPNe  NGC~40 [WC8] and BD+30$^{\circ}$ 3639 [WC9] using distances and $c(H\beta)$ from \citet{BucciarelliStanghellini2023}.  Each have $M_{J} \sim +0.5$ mag using 2MASS photometry of  their central stars. Similar arguments apply to the more distant, also late-type, [WC] sources WR-C-13 (G307.2--0.6) and WR-C-45 (G31.2--1.3) from \citet{Mulato+2025}. We marginally favour a massive star nature for WR84-13 (WC9d) despite its modest absolute J-band magnitude owing to its location in the thin-disk ($z$ = 70 pc) and its estimated bolometric luminosity ($\log L/L_{\odot} \sim 4.5$) exceeds that of luminous post-AGB stars \citep{Blocker1995}.


Of the {\it Gaia} ESP-ELS candidates observed by \citet{Mulato+2025}, WR-C-06 reveals  emission at He\,{\sc i} 1.08$\mu$m and P$\alpha$, although there is no evidence for He\,{\sc ii} emission, in agreement with.  their
conclusions that this is not a WR star. Several additional candidates remain viable WR stars on the basis of their spectrophotometry with SPHEREx, including the candidate WN star WR-C-53 for 
which He\,{\sc ii} 1.01$\mu$m, 4.76$\mu$m are detected with low significance.  

Another  candidate which will be denoted by its RA/Dec coordinates (J1823+0250) is confirmed as a (weak) He emission line star, including He\,{\sc ii} 1.01$\mu$m, 1.86$\mu$m, 3.09$\mu$m, 4.76$\mu$m, plus He\,{\sc i} 1.08$\mu$m 
(see Fig.~\ref{He-emission} in the Appendix).  We estimate its J-band extinction from its $G_{\rm BP} - G_{\rm RP}$  {\it Gaia} colour \citep{RateCrowther2020}, indicating an absolute J-band magnitude of $M_{\rm J} =$ --0.1$^{+0.4}_{-0.4}$ mag,  
several magnitudes fainter than WN stars \citep{RateCrowther2020}, and its {\it Gaia} parallax infers a large distance from the 
galactic plane ($|z| = 0.78^{+0.16}_{-0.12}$ kpc). Consequently J1823+0250 is a potential intermediate mass (stripped) Helium star, for which \citet{Gotberg+2018} solar 
metallicity He star models with 4.5 to 5.1 $M_{\odot}$ infer absolute J-band magnitudes of +0.3 and --0.45 mag, respectively.  
Indeed, absolute J-band magnitudes of stripped He stars in the Magellanic Clouds \citep{Drout+2023, Gotberg+2023} range from --1.0 mag (star 1, star 5) to +0.9 mag (star~4) 
on the basis of VMC photometry \citep{VMC}. 

Intermediate mass stripped He stars are generally expected to possess weak winds \citep{SanderVink2020}. J1823+0250 merits further study since it may correspond to a transition 
spectrum  between a dominant WR-like emission at high mass, and subdwarf O-type absorption at low mass \citep[][their figure~5]{Gotberg+2018}.
In general the low resolving power hinders an assessment of whether He\,{\sc ii} emission is stellar or nebular, but the higher resolving power of SPHEREx Band B6 is consistent either with a 
nebular origin of He\,{\sc ii} 4.76$\mu$m, or a slow wind if it is stellar in origin, indicating of a hot, He-rich source in either case. 

Sources with spectroscopic similarities to WR stars in the visual include Cataclysmic Variables (CVs) - indeed \citet{Chang+2025} attributed a spectrum of the CV SW Sex from LAMOST DR10 to a WN star. Excluding any foreground J-band extinction, the absolute J-band magnitude of SW~Sex is +5.3 mag based on its {\it Gaia} DR3 parallax-derived  distance of 0.60 kpc \citep{bailer-jones2021}, so one can exclude such an origin for J1823+0250. 

\citet{Muller-Horn2026+} have recently identified a WN star in a close binary system, WR2-1, from SDSS-V Milky Way Mapper (MWM) spectroscopy \citep{SDSS-V}.  Strong emission-lines from the WN star are 
heavily diluted by the brighter O star companion at visual wavelengths, which may explain why it eluded detection from previous H$\alpha$ surveys. 
WR2-1 has a systemic absolute magnitude of $M_{\rm J}$ = --4.1$^{+0.2}_{-0.2}$ mag assuming a {\it Gaia} DR3 photogeometric distance of 6.45$^{+0.56}_{-0.66}$~kpc \citep{bailer-jones2021}. Adopting the extinction of $A_{\rm V} = 5.5$ mag ($A_{\rm J} = 1.5$ mag) and the 25$\pm$5\% contribution of the WR star
to the J-band flux from \citet{Muller-Horn2026+}, its absolute J-band magnitude is --2.4$^{+0.3}_{-0.3}$ mag, comparable to many newly identified WN stars from this study. SPHEREx spectroscopy of WR2-1 is included in the Appendix (Fig.~\ref{He-emission}), with He\,{\sc ii} 3.09$\mu$m, 4.76$\mu$m emission lines observed, plus P$\alpha$+He\,{\sc ii} and Br$\alpha$+He\,{\sc ii}. He\,{\sc ii} 1.01$\mu$m is not observed, presumably 
owing to the greater contribution from the O star companion at shorter wavelengths, He\,{\sc ii} 4.76$\mu$m has a modest FWHM owing to the slow wind of its WN component. 


\begin{table}
    \caption{Overview of key  stellar and nebular emission line diagnostics observed in the SPHEREx wavelength range for WNE, WNL, WCE/WO (and [WCE]/[WO]), WCL (and [WCL]), Planetary Nebulae (PNe), O and B supergiants.
    Parenthesis indicate weak feature.}
    \centering
    \label{summary}
    \begin{tabular}{
l@{\hspace{0mm}}
l@{\hspace{0mm}}
c@{\hspace{1.5mm}}
c@{\hspace{1.5mm}}
c@{\hspace{1.5mm}}
c@{\hspace{1.5mm}}
c@{\hspace{1.5mm}}
c@{\hspace{1.5mm}}
c}
    \hline
    $\lambda$  & Ion                                                              & WNE             & WNL           & WCE/         & WCL          & PNe             & O        & B  \\
    ($\mu$m)    &                                                                   &                      &                    & WO            &                   &                     & \multicolumn{2}{c}{supergiant} \\
 \hline
    0.77                          & C\,{\sc iv} 7-6                            & --                 & --                    & \Checkmark & --                     & --                 & --                 & -- \\
    0.95                          & [S\,{\sc iii}]                                   & --                  & --                  & --                  & --                  & \Checkmark  & --                 & -- \\
    0.97                          & C\,{\sc iii}  3d-3p                        & --                  &                     &(\Checkmark) & \Checkmark  & --                  & --                 &  -- \\
    1.01                          & He\,{\sc ii} 5-4+C\,{\sc iv} 10-8                        & \Checkmark & (\Checkmark) & \Checkmark & (\Checkmark) & (\Checkmark) & (\Checkmark) & --  \\  
    1.08                          & He\,{\sc i} 2p-2s                             & (\Checkmark) & \Checkmark & (\Checkmark)  & \Checkmark &  \Checkmark & (\Checkmark)  & \Checkmark\\
    1.19                          & C\,{\sc iv}  8-7                         & --                 & --                    & \Checkmark &  --                  & --                 & --                 & -- \\
    1.28                          & P$\beta$                                     &(\Checkmark)  & (\Checkmark) & --                & --                       & \Checkmark & --                 & (\Checkmark) \\
    1.74                          & C\,{\sc iv}  9-8                         & --                 & --                    & \Checkmark &--                    & --                 & --                 & -- \\
    1.87                          & P$\alpha$+He\,{\sc ii}  6-5 & \Checkmark & \Checkmark & \Checkmark & (\Checkmark) & \Checkmark  & (\Checkmark) & \Checkmark \\
     2.06                          & He\,{\sc i}  2p-2s                           & --                   & (\Checkmark) & --                 & (\Checkmark) & --                & --                & (\Checkmark) \\
    2.08                           & C\,{\sc iv} 3d-3p                              & --                  & --                  &  \Checkmark &( \Checkmark)  & --                 & --                  & -- \\
    2.43                           & C\,{\sc iv} 10-9                         & --                  & --                  & \Checkmark & --                      & --                 & --                  & -- \\
    2.62                           & Br$\beta$                                   &(\Checkmark) & (\Checkmark) & --                & --                       & \Checkmark & --                 & (\Checkmark) \\
    3.09                           & He\,{\sc ii} 7-6+C\,{\sc iv} 14-12                         & \Checkmark & (\Checkmark) & \Checkmark & (\Checkmark) & (\Checkmark) & (\Checkmark) & -- \\
    3.28                           & C\,{\sc iv}  11-10                       & --                 & --                    & \Checkmark & --                     & --                 & --                 & -- \\
    4.05                           & Br$\alpha$+He\,{\sc ii} 10-8     & \Checkmark & \Checkmark & \Checkmark & (\Checkmark) & \Checkmark  &( \Checkmark) & \Checkmark \\
    4.30                           & C\,{\sc iv} 12-11+He\,{\sc i} 3p-3s     & (\Checkmark)  & \Checkmark & \Checkmark  & (\Checkmark) &  --                 & --                 & \Checkmark \\ 
    4.49                            & [Mg\,{\sc iv}]                                & --                  & --                  & --                  & --                   & \Checkmark  & --                 & --           \\
      4.65                          & Pf$\beta$                                   & (\Checkmark)   & (\Checkmark) & --                & --                       & \Checkmark & --                 & (\Checkmark) \\
4.76                            & He\,{\sc ii} 8-7+C\,{\sc iv} 16-14  &  \Checkmark & (\Checkmark) & \Checkmark & (\Checkmark) & (\Checkmark)& (\Checkmark) & -- \\
     \hline
    \end{tabular}
    \end{table}

\subsection{Low probability candidates}

%

We have  extracted SPHEREx observations for  8 additional ESP-ELS candidate WR
stars with {\tt classlabel$\_$espels$\_$flag} = 3 or 4. Most candidates are at relatively high Galactic
latitude so are unlikely to be genuine Milky Way WR stars, so we  focus solely on sources with $|b| \leq 10^{\circ}$. These include WR-C-34  \citep{Mulato+2025} which remains a viable
 WC candidate since C\,{\sc iii} 0.97$\mu$m is detected, but the spectrum is unusable at longer wavelengths. WR-C-04 exhibits strong P$\alpha$ and Br$\alpha$ emission, although lacks any evidence for He\,{\sc i-ii} emission. Br$\alpha$, Pf$\beta$ emission is also observed in WR-C-02, while the only prominent spectral feature in WR-C-07 is 3.3$\mu$m PAH emission. The lack of He\,{\sc ii} emission in WR-C-02, WR-C-04 or WR-C-07 agrees with the non-WR identification of these sources by \citet{Mulato+2025}.

 Another He emission line star amongst the ESP-ELS candidates is J1907-0523, whose SPHEREx spectral morphology is consistent with an early-type, weak-lined WN star (Fig.~\ref{He-emission}). 
 However, a massive star origin for J1907-0523 is excluded (as for J1823+0250) owing to its modest absolute J-band magnitude of $M_{\rm J}$ = +0.4$^{+0.4}_{-0.4}$ mag on the basis of a {\it Gaia} DR3 
 photogeometric distance of 7.1$^{+1.6}_{-1.4}$ kpc \citep{bailer-jones2021}, and $A_{\rm J}$ estimated from its {\it Gaia}  $G_{\rm BP} - G_{\rm RP}$ colour. This source also lies at a  large distance from the mid-plane ($|z| = 0.75^{+0.15}_{-0.15}$ kpc), and
 matches the predicted J-band absolute magnitude of a $\sim$4$M_{\odot}$ stripped He star model \citep{Gotberg+2018}.  In contrast with J1823+0250, the higher resolving power of SPHEREx in Band B6 is indicative of a stellar origin of He\,{\sc ii} 4.76$\mu$m. Consequently, it is a second potential intermediate mass He star,  so merits further study.  
 Alternatively, it might be belong to the rare subset of PNe with [WN]-type central stars \citep[e.g. IC~4663][]{Miszalski+2012}, but J1907-0523 lacks the strong nebular emission usually characteristic of PNe (NGC~2392 in Fig.~\ref{CSPNe}).

Three low probability WR candidates at high Galactic latitude are established [WC]-type PN central stars, NGC~6751 \citep{Crowther+1998}, RaMul~2 \citep{Werner+2024} and PMR~1 \citep{Morgan+2001}. Of these NGC~6751 and PMR~1 exhibit prominent emission lines in SPHEREx spectrophotometry, including broad 3.3$\mu$m PAH emission, while RaMul~2 is faint, with solely prominent C\,{\sc iv} 2.08$\mu$m, 4.30$\mu$m emission. SPHEREx spectrophotometry for NGC~6751 is included in Fig.~\ref{cspn_obsuper}, with the [S\,{\sc iii}] 0.91+0.95$\mu$m doublet observed, while PMR~1 is presented in the Appendix (Fig.~\ref{CSPNe}).

\begin{figure}
    \centering
     \includegraphics[width=\columnwidth,angle=0,bb=50 72 542 761]{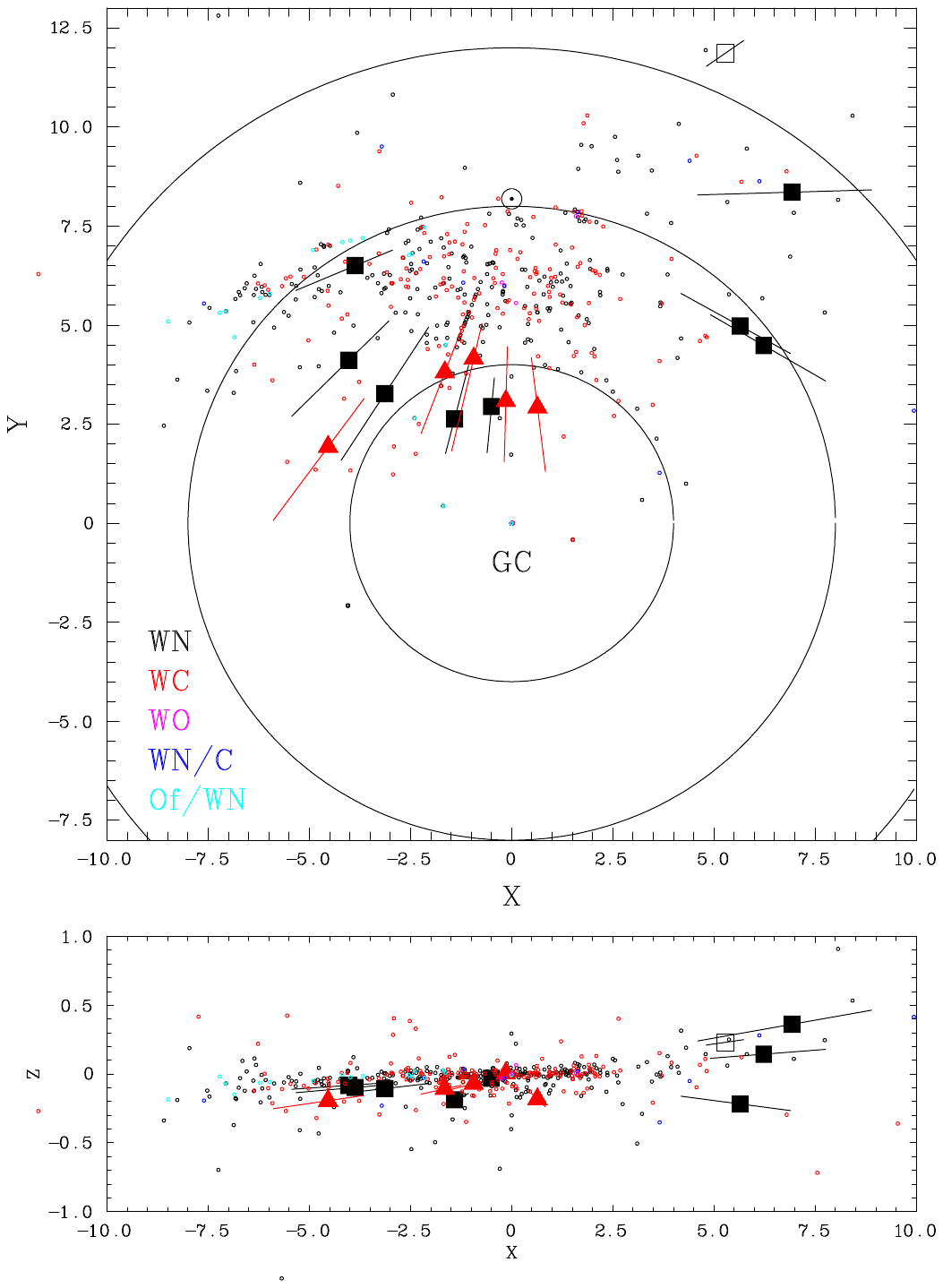}
         \caption{Location of newly identified WR stars - black squares for WN stars, red triangles for WC stars - together with estimated distances to previously known WR stars (open circles, key indicates subtypes) on a top down ($X, Y$) and side view ($X, Z$) of the Milky Way, in which the Galactic Centre is located at $X=Y=Z$=0 and the Sun is located at $X$=0, $Y$=8.15 kpc \citep{EHT}. Galactocentric distances of 4, 8 and 12 kpc are indicated, as is the recently identified
         O+WN system WR2-1 \citep[open square][]{Muller-Horn2026+}}
         \label{gal_structure}
    \end{figure}

\section{Discussion and conclusions}\label{conc}

Up until the advent  of (very) large spectrophotometric datasets, courtesy of {\it Gaia} XP and SPHEREx, establishing the nature of Galactic emission line stars was a  laborious, resource intensive undertaking. Candidates  identified in photometric surveys, using narrow-band filters required dedicated spectroscopic follow-up at optical or IR wavelengths. \citet{Marin+2024} and \citet{Mulato+2025} have recently exploited the ESP-ELS analysis \citep{Creevey+2023} of {\it Gaia} XP spectrophotometry to confirm WR candidates using higher spectral resolution observations from ground-based facilities. High foreground
dust extinction and the low resolving power hinders the use of {\it Gaia} XP in isolation since the primary spectral diagnostic (He\,{\sc ii} 0.469$\mu$m) lies in the blue visual.

We have used additional candidate Wolf-Rayet stars identified from ESP-ELS \citep{Creevey+2023} to illustrate the potential use of SPHEREx for the discovery of previously unknown WR and related stars, since useful diagnostics span the entire
0.75--5.0$\mu$m wavelength range. Care needs to be taken to exclude other emission line sources since the low resolving power of SPHEREx usually prevents strong nebular emission being distinguished from stellar emission. Table~\ref{summary} provides an overview of key stellar and nebular emission lines observed in the SPHEREx wavelength range for WR, PNe and OB supergiants. Parentheses
indicate weak emission features observed for each source type. SPHEREx is sufficiently sensitive that relatively isolated WR stars in the Magellanic Clouds can be easily identified, such as the WO binary Sk~188 in the SMC (Fig.~\ref{SPHEREx-WR}).

Historical searches using He\,{\sc ii} 0.469$\mu$m emission and the adjacent continuum are sensitive to both WN and WC stars \citep{Shara+1999}. Subsequent narrow-band IR photometric surveys conducted from ground-based facilities have generally focused on the K$_{s}$-band \citep{Crowther+2006, Shara+2009} in order to avoid high dust extinction at shorter wavelengths. Unfortunately, multiple filters are required in the K$_{s}$-band, since (most) WC stars exhibit strong C\,{\sc iv} 2.08$\mu$m emission, while (most) WN stars show weaker emission in either He\,{\sc ii} 2.19$\mu$m or Br$\gamma$, such that \citet{Shara+2009} adopted a six filter system. Since SPHEREx provides access to the full 0.75--5$\mu$m wavelength range, He\,{\sc ii} 3.09$\mu$m (mindful of potential PAH emission) or He\,{\sc ii} 4.76$\mu$m emission potentially provide spectral diagnostics of additional highly reddened WR stars, since only very late-type and dusty WR stars are weak in these lines.

Fig.~\ref{gal_structure} illustrates the location of the new WR stars within the Milky Way (top down, and side view), together with locations of all previously confirmed WR stars on the basis of  distances from {\it Gaia} DR3 parallaxes \citep{RateCrowther2020, bailer-jones2021}, the distance to the Galactic Centre \citep{EHT} , distances to star clusters \citep{Rate+2020}, or distance estimates from absolute magnitude calibrations \citep{RossloweCrowther2015} otherwise. It is apparent that the newly reported WR stars are also on the near side of the Milky Way, with an average distance of 5.7$\pm$1.1 kpc. This is unsurprising since candidates are drawn from {\it Gaia} XP spectroscopy which is limited to sources brighter than $m_{G}$ = 19 mag. Most WR stars are close to the mid-plane, although some runaways are observed at surprisingly high vertical distances \citep{RateCrowther2020, Arthur2025}. The recently discovered O+WN system WR2-1 is indicated in Fig.~\ref{gal_structure} with a large open square.

Representative WR stars on the far side of the Milky Way, at galacto-centric distances of 4--8 kpc will lie 9--16 kpc away from the Sun (distance moduli 14.8 to 16.0), with a typical K$_{s}$-band dust extinction of $A_{\rm K_{s}} \sim$ 1.5 mag (higher in the direction of the Galactic Centre). Absolute K$_{s}$-band magnitudes of (non dusty) WR stars range from $M_{\rm K_{s}}$ = --2 to --7 mag, but --5 mag is typical of mid WN and WC stars \citep{RossloweCrowther2018}, so anticipated K$_{s}$-band magnitudes are 11.3 $\leq$ K$_{s} \leq$ 12.5 mag, well within the capabilities of SPHEREx. 

Of course the faintest WR stars towards the dustiest sight-lines ($A_{\rm K_{s}}$ = 3 mag), with K$_{s} \sim$ 17 mag, would be challenging even neglecting the coarse spatial scale of SPHEREx. Distances to such objects would require infrared astrometry with the proposed {\it GaiaNIR} mission \citep{GaiaNIR}. Nevertheless, SPHEREx  offers the prospect of  increasing the known WR content of the Milky Way from targeted searches. All newly confirmed WR stars satisfy the primary IR photometric colour criteria of 
\citet{Faherty+2014}, as do most stellar dominated [WC]-type stars (the only exception is PMR 1).

In addition to searches for new WR stars, SPHEREx also permits previously reported WR stars to be verified. By way of example,
the coordinates of the WN star 1603-75L (= WR123-4) from \citet[][]{Shara+2012} correspond to a non-2MASS source, yet they provide photometry for a 2MASS source 11 arcsec to the north with K$_{s}$=13.68 mag. 
SPHEREx spectrophotometry of WR123-4 shown in Fig.~\ref{CSPNe}  confirms a late WC star at their quoted position, with K$_{s}\sim$12.4 mag.

In summary, we have presented a test application of SPHEREx for the discovery of Galactic WR stars, involving candidate WR stars using {\it Gaia} XP. Excluding recently confirmed candidates \citep{Marin+2024, Mulato+2025}, we confirm an additional 13 Milky Way WR stars, 8 WN and 5 WC, discuss similarities with [WC]-type CSPNe, and identify a few He emission line stars far from the galactic plane which may be intermediate mass He stars. The sensitivity of SPHEREx is capable of detecting WR stars on the far side of the Milky Way, indeed one example of a WR system in the SMC is presented. Consequently, SPHEREx has the potential to detect many more WR stars, by combining Machine Learning techniques with a judicious choice of candidate selection criteria.


\section*{Acknowledgements}

This research was made possible thanks to a Sheffield University Research Experience (SURE) summer placement for ELS. Thanks to Bill Vacca for providing the IRTF/SpeX datasets for a sample of Galactic Wolf-Rayet stars, and Lionel Mulato for sharing optical spectroscopy of a subset of ESP-ELS candidates.  Thanks also to Ylva G\"{o}tberg for helpful discussions regarding stripped Helium stars during the August 2026 Munich Institute for Astro-, Particle and BioPhysics (MIAPbP) workshop. MIAPbP  is funded by the Deutsche Forschungsgemeinschaft (DFG, German Research Foundation) under Germany´s Excellence Strategy – EXC-2094 – 390783311. Comments from the anonymous referee helped clarify several aspects of the original manuscript.

This publication makes use of data products from the Spectro-Photometer for the History of the Universe, Epoch of Reionization and Ices Explorer (SPHEREx), which is a joint project of the Jet Propulsion Laboratory and the California Institute of Technology, and is funded by the National Aeronautics and Space Administration. This research has also made extensive use of the NASA/IPAC Infrared Science Archive, which is funded by the National Aeronautics and Space Administration and operated by the California Institute of Technology. This publication makes use of data products from the Wide-field Infrared Survey Explorer, which is a joint project of the University of California, Los Angeles, and the Jet Propulsion Laboratory/California Institute of Technology, funded by the National Aeronautics and Space Administration. This research has made (extensive) use of the SIMBAD database, operated at CDS, Strasbourg, France.

\section*{Data Availability}

Photometric IR catalogues from 2MASS, {\it Spitzer}, {\it WISE} are available at the NASA/IPAC Infrared Science Archive (IRSA)\footnote{\url{https://irsa.ipac.caltech.edu/applications/Gator/}} while SPHEREx spectrophotometry can also be accessed from IRSA\footnote{\url {https://irsa.ipac.caltech.edu/applications/spherex/tool-spectrophotometry}}



\bibliographystyle{mnras}
\bibliography{spherex} 




\appendix

\section{SPHEREx spectrophotometry of additional sources}

\begin{figure*}
    \centering
     \includegraphics[width=0.9\columnwidth,angle=0]{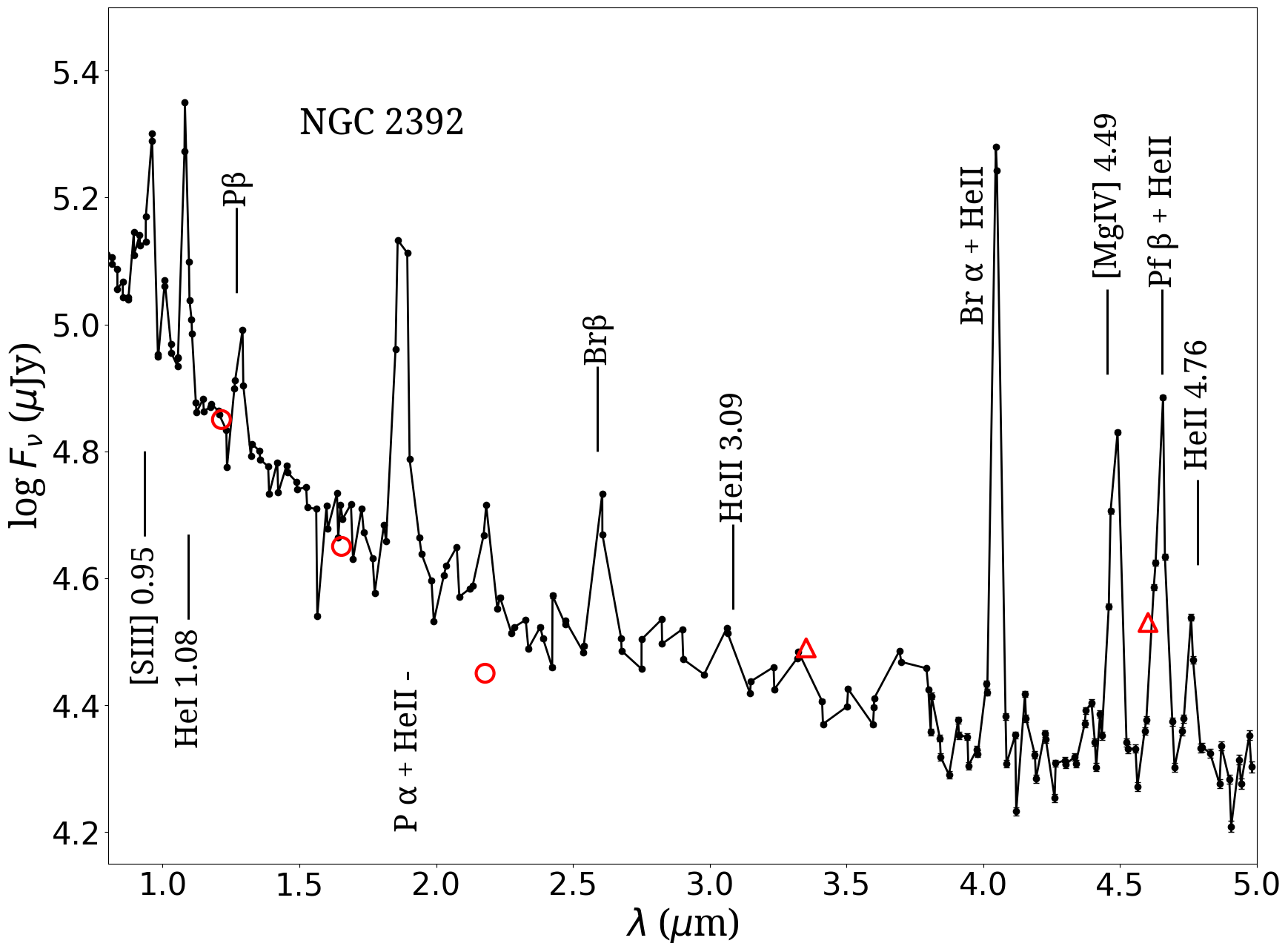}
     \includegraphics[width=0.9\columnwidth,angle=0]{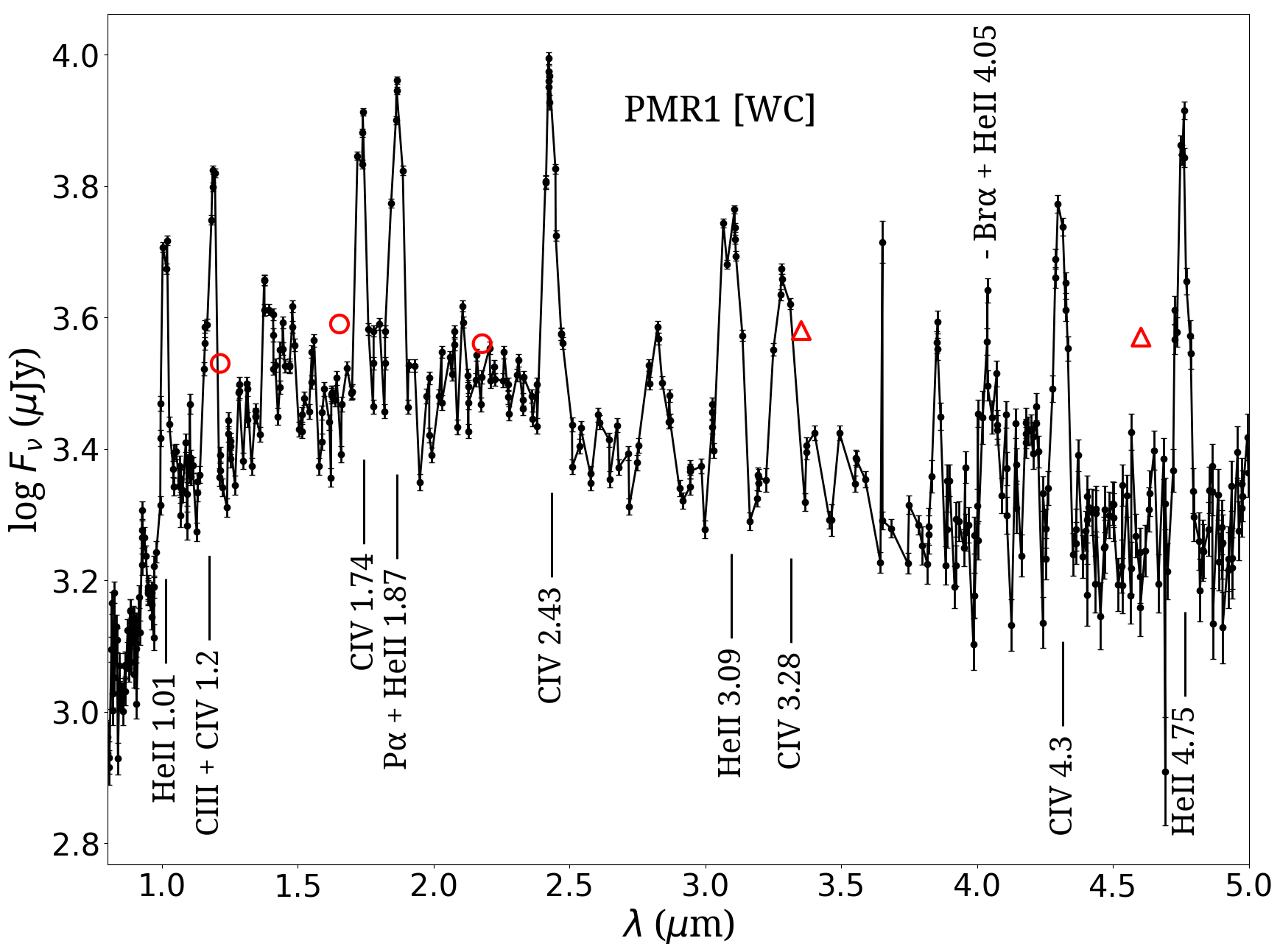}
          \includegraphics[width=0.9\columnwidth]{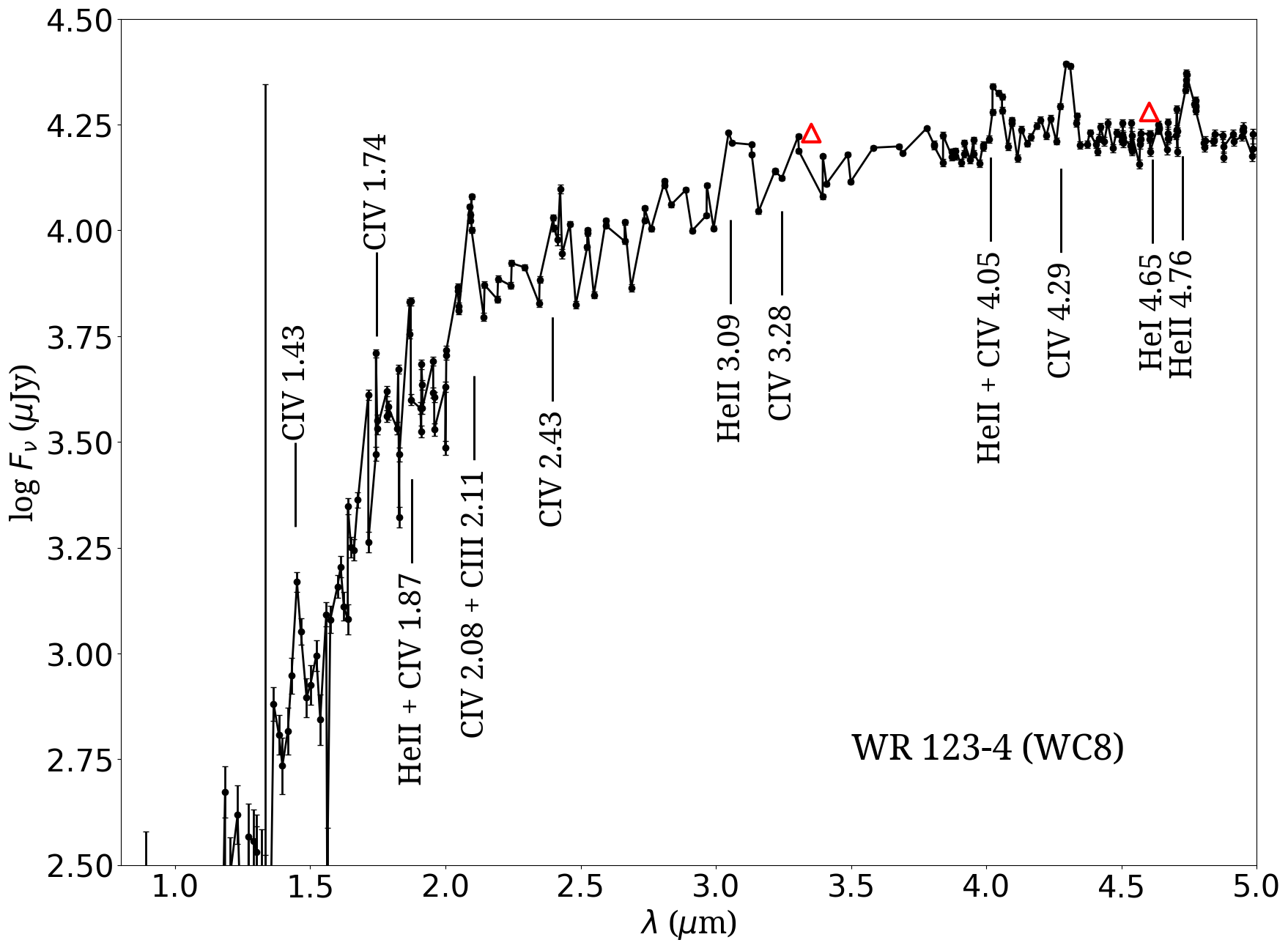}
    \includegraphics[width=0.9\columnwidth,angle=0]{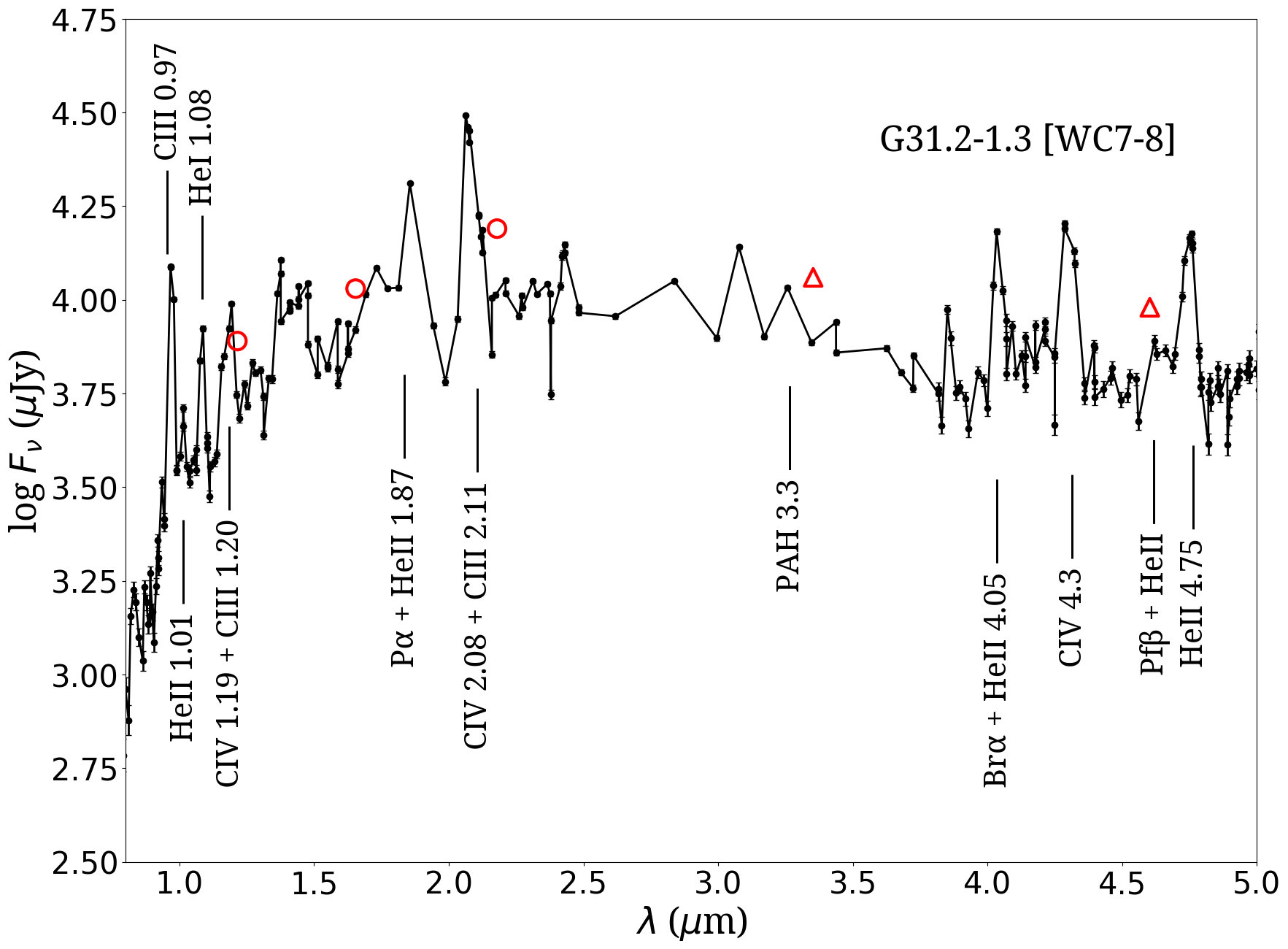}
    \includegraphics[width=0.9\columnwidth,angle=0]{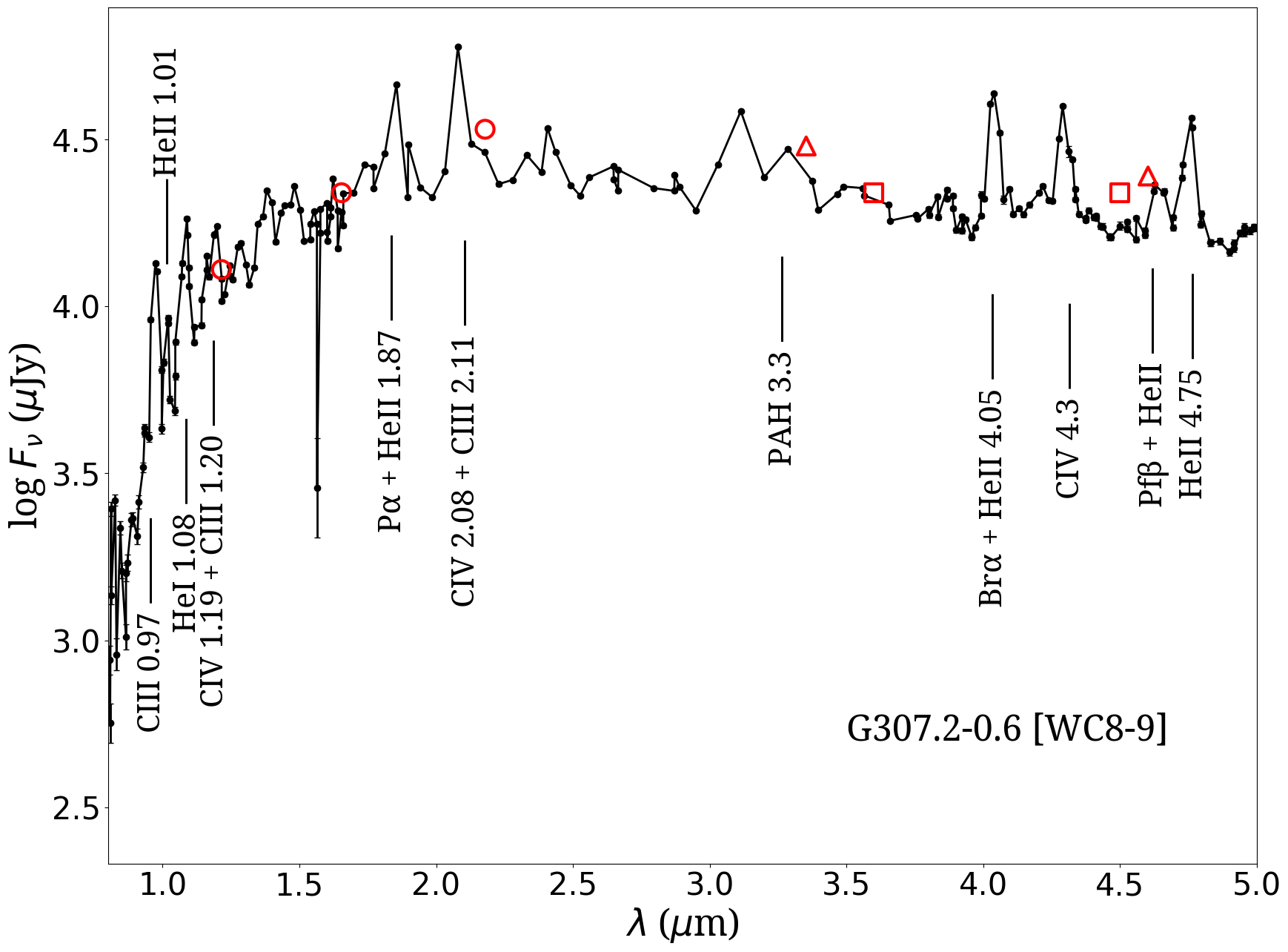}
     \includegraphics[width=0.9\columnwidth,angle=0]{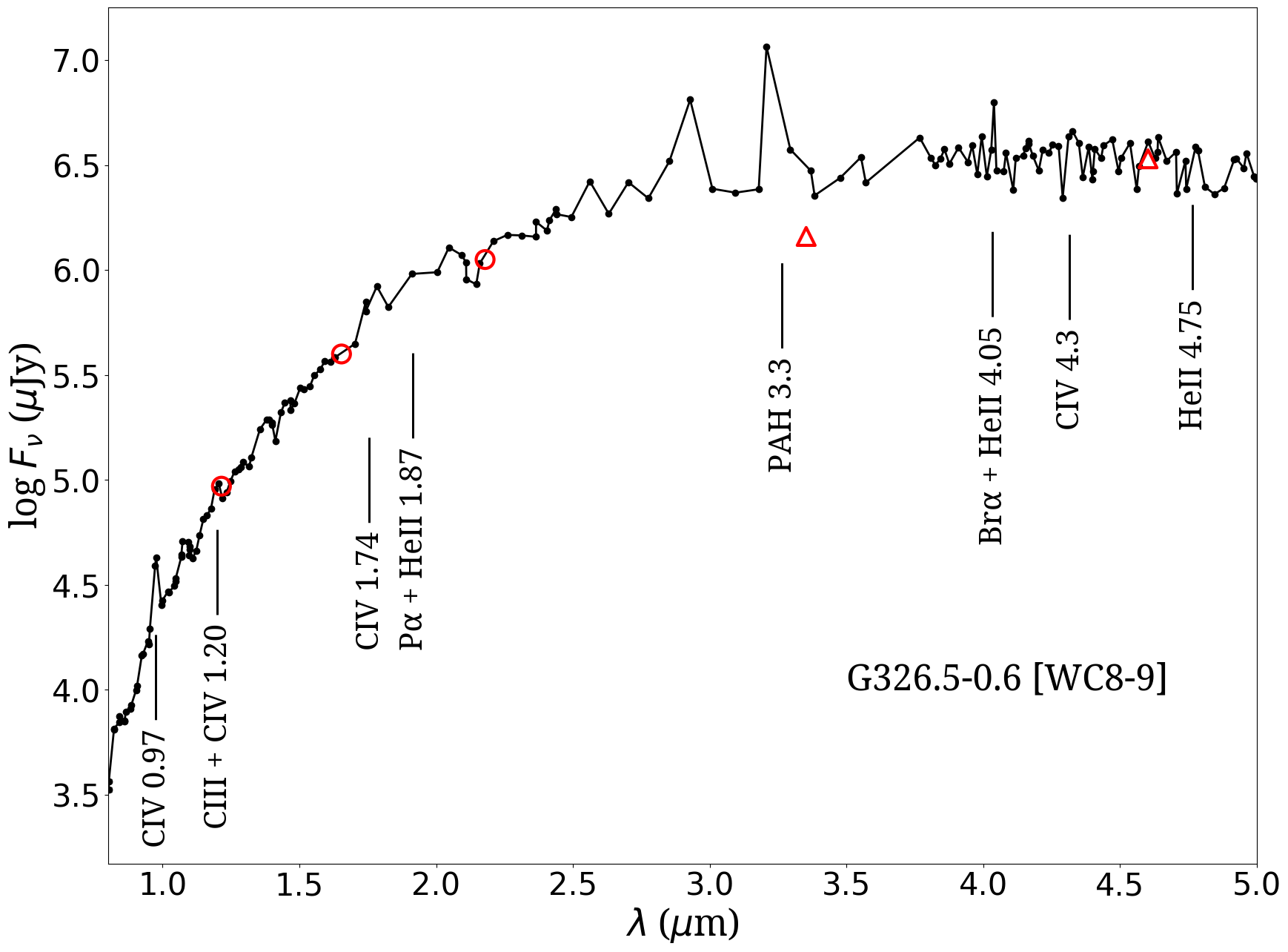}
    \caption{SPHEREx spectrophotometry of the Planetary Nebula NGC~2392 (top left), [WC]-type central star PMR~1 (top right), WC8 star WR123-4 (middle left) a.k.a. 1603-75L from \citet{Shara+2012}, 
    plus newly identified sources G31.2-1.3 (middle right), G307.2-0.6  (bottom left) and G326.5-0.6 (bottom right) together with photometry from 2MASS (open circles), 
    {\it Spitzer}/GLIMPSE (open squares) and {\it WISE} (open triangles). Spectrophotometry of WR123-4 infers near-IR photometry of J$\sim$16.7 mag, H$\sim$14.3 mag and K$\sim$12.4, 
    contrary to the 2MASS source 11$''$ to its north with K=13.68 mag quoted by \citet{Shara+2012}.}
         \label{CSPNe}
    \end{figure*}

Figure~\ref{CSPNe} presents SPHEREx spectrophotometry of the Planetary Nebula NGC~2392, [WC4]-type central star PMR 1 \citep{Morgan+2001},  WC8 star WR123-4 a.k.a. 1603-75L from \citet{Shara+2012}, 
plus newly identified sources G31.2-1.3 [WC7-8] G307.2-0.6 [WC8-9], G326.5-0.6 [WC9] together with  photometry from 2MASS (open circles), {\it Spitzer}/ GLIMPSE (open squares) and {\it WISE} (open triangles). 
Nebular emission dominates the 1--5$\mu$m spectrum of NGC~2392, whereas stellar wind He\,{\sc ii} and C\,{\sc iv} lines dominate the spectrum of PMR~1, with C\,{\sc iv} 2.08$\mu$m weak, 
as in the [WO]-type central star NGC~6905 and [WC]-type central star NGC~6751 (Fig.~\ref{cspn_obsuper}). WR123-4 exhibits the blend of C\,{\sc iv} 2.08$\mu$m with C\,{\sc iii} 2.11$\mu$m, but suffers from high
dust extinction, with J$\sim$16.7 mag. The candidate CSPNe host late spectral type central stars, so also include C\,{\sc iii} 0.97$\mu$m, with C\,{\sc iv} 2.08$\mu$m prominent in G31.2-1.3 and G307.2-0.6.


\begin{figure}
    \centering
     \includegraphics[width=0.95\columnwidth,angle=0]{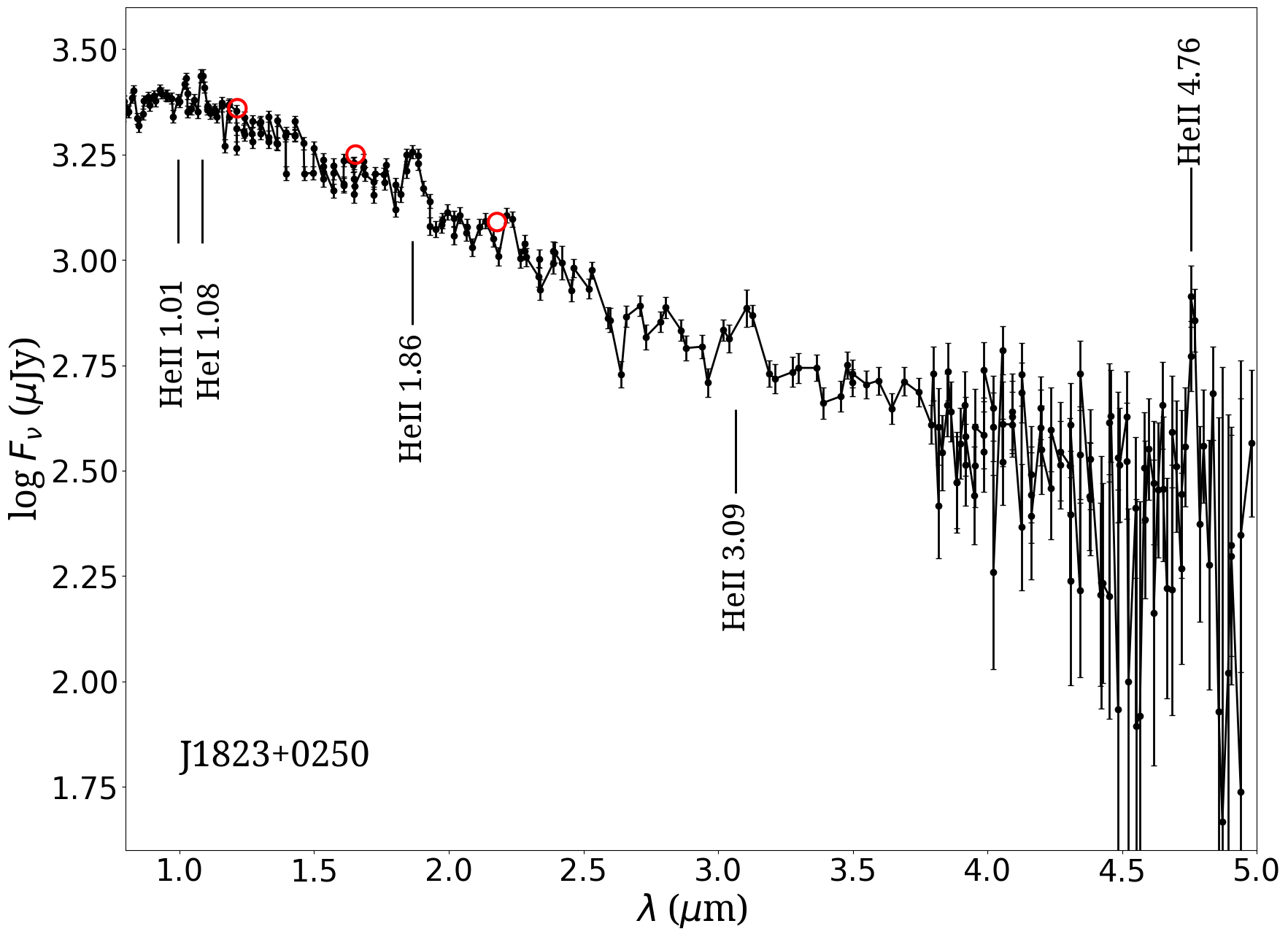}
    \includegraphics[width=0.95\columnwidth,angle=0]{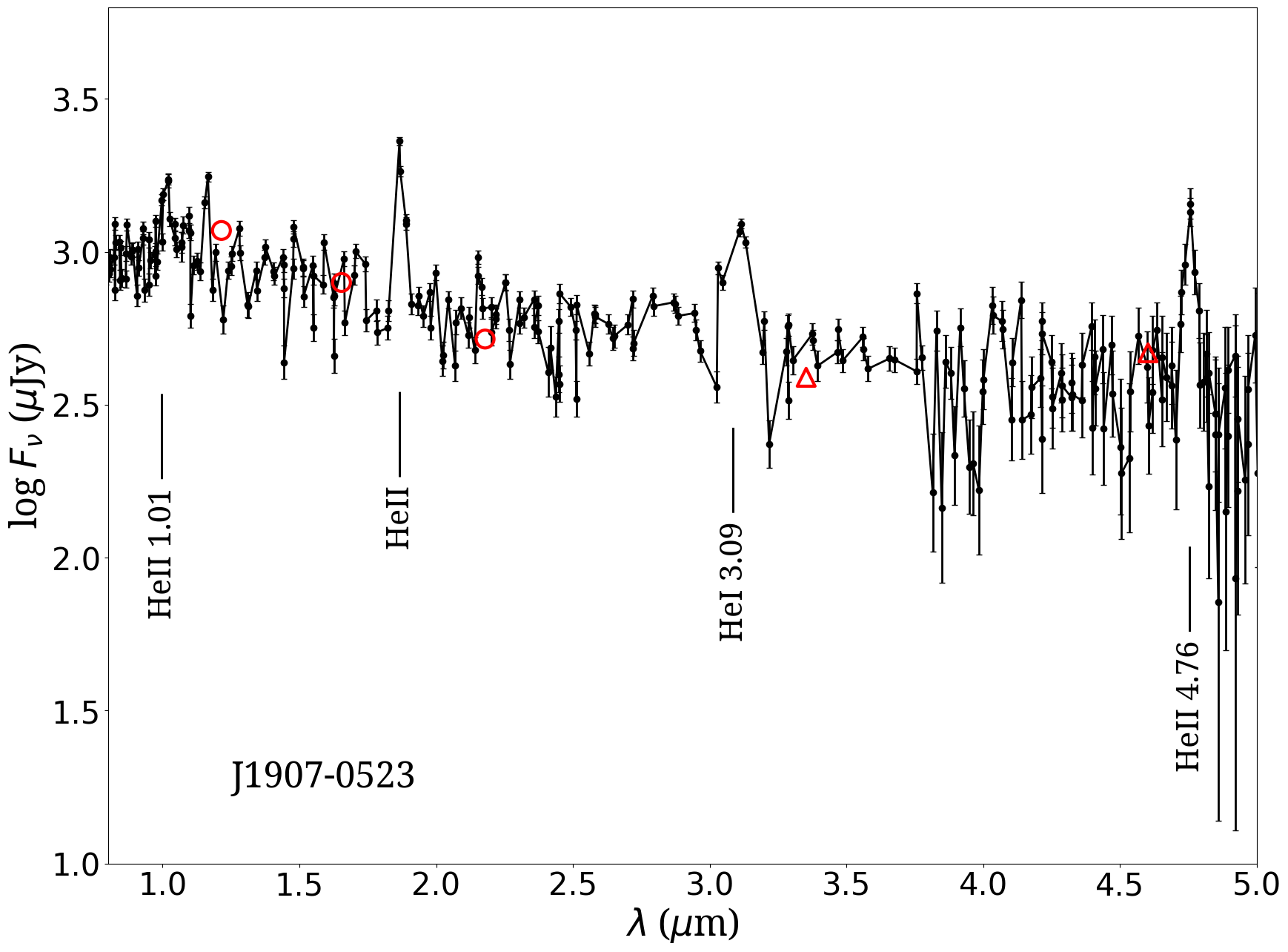}
    \includegraphics[width=0.95\columnwidth,angle=0]{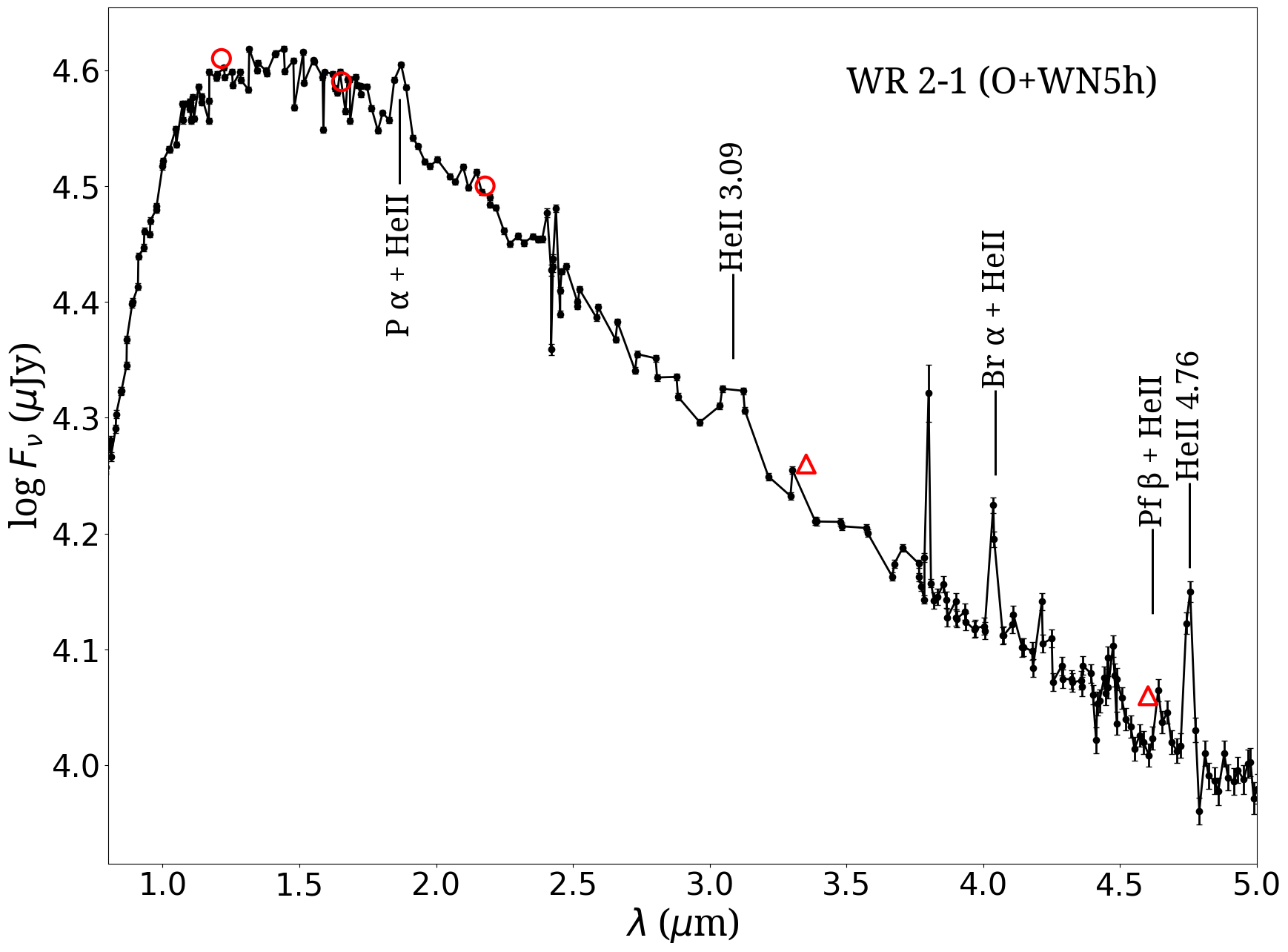}
    \caption{SPHEREx spectrophotometry of the He\,{\sc ii} emission line sources J1823+0250 (top) and J1907-0523 (middle), plus the newly identified O+WN system WR2-1 from \citet[][bottom]{Muller-Horn2026+}, together with photometry from 2MASS (open circles) and {\it WISE} (open triangles).}
         \label{He-emission}
    \end{figure}

Figure~\ref{He-emission} presents SPHEREx spectrophotometry of the  He \,{\sc ii} emission line sources 
J1823+0250 and J1907-0523, and the recently identified WN binary WR2-1 \citep{Muller-Horn2026+}, together with  photometry from 2MASS (open circles) and {\it WISE} (open triangles). 
J1823+0250 and J1907-0523 both show  He\,{\sc ii} 1.01$\mu$m,  1.86$\mu$m, 3.09$\mu$m,  4.76$\mu$m, with He\,{\sc i} 1.08$\mu$m also observed in J1823+0250. 
The higher resolving power of SPHEREx in Band B6 perhaps an assessment of stellar vs nebular origin of He\,{\sc ii} 4.76$\mu$m. This is consistent with a nebular in origin for J1823+0250, whereas J1907-0523 is significantly broadened, suggesting a stellar origin.
WR2-1 reveals prominent He\,{\sc ii} 3.09$\mu$m and 4.76$\mu$m, plus P$\alpha$+He\,{\sc ii} and Br$\alpha$+He\,{\sc ii}.  Notably, the He\,{\sc ii} 4.76$\mu$m emission in WR2-1 could also be inferred to be nebular in origin owing to its slow wind \citep{Muller-Horn2026+}.

\bsp
\label{lastpage}
\end{document}